# Strong coupling between coherent ferrons and cavity acoustic phonons

Yujie Zhu[1], Jiaxuan Wu[1], Anna N. Morozovska[2], Eugene A. Eliseev[3], Yulian M. Vysochanskii[4], Venkatraman Gopalan[5], Long-Qing Chen[5], Xufeng Zhang[6,8], Wei Zhang[7], Jia-Mian Hu[1*]

[1] *Department of Materials Science and Engineering, University of Wisconsin-Madison, Madison, WI, 53706, USA*

[2] *Institute of Physics, National Academy of Sciences of Ukraine, 46, pr. Nauky, 03028 Kyiv, Ukraine*

[3] *Frantsevich Institute for Problems in Materials Science, National Academy of Sciences of Ukraine, Omeliana Pritsaka str., 3, Kyiv, 03142, Ukraine*

[4] *Institute of Solid-State Physics and Chemistry, Uzhhorod University, 88000 Uzhhorod, Ukraine*

[5] *Department of Materials Science and Engineering, The Pennsylvania State University, University Park, PA 16802, USA*

[6] *Department of Electrical and Computer Engineering, Northeastern University, Boston, Massachusetts 02115, USA*

[7] *Department of Physics and Astronomy, University of North Carolina at Chapel Hill, Chapel Hill, NC 27599, USA*

[8] *Department of Physics, Northeastern University, Boston, Massachusetts 02115, USA*

## Abstract

Coherent ferrons, the quanta of polarization waves, can potentially be hybridized with many other quasiparticles for achieving novel control modalities in quantum communication, computing, and sensing. Here, we theoretically demonstrate a new hybridized state resulting from the strong coupling between fundamental-mode (wavenumber is zero) coherent ferrons and cavity bulk acoustic phonons. Using a van der Waals ferroelectric $CuInP_2S_6$ membrane as an example, we predict an ultra-strong ferron-phonon coupling at room temperature, where the coupling strength $g_c$ reaches over 10% of the resonant frequency $\omega_0$. We also predict an in-situ bistable electric-field control of mode-specific ferron-phonon hybridization via ferroelectric switching. We further show that $CuInP_2S_6$ allows for reaching the fundamentally intriguing but challenging deep-strong coupling regime (i.e., $g_c/\omega_0$>1) near the ferroelectric-to-paraelectric phase transition. Our findings establish the theoretical basis for exploiting coherent ferron as a new contender for hybrid quantum system with strong and highly tunable coherent coupling.

*E-mail: jhu238@wisc.edu

***Introduction.*** Hybridization between elementary excitations in different physical systems leads to the creation of new coherent states, with potential applications in quantum communication, computing, and sensing [1,2]. Such hybridization is typically characterized by the coupling strength $g_c$, which determines the rate of energy exchange. The strong coupling regime, when the coupling exceeds the respective energy dissipation rates of each system $\kappa_1$ and $\kappa_2$ (i.e., $g_c/\kappa_1>1$ and $g_c/\kappa_2>1$), is a desirable condition for quantum transduction [3]. For example, strong coupling between magnons (elementary excitation of magnetization) and microwave photons [4,5] has enabled a magnon-photon-qubit transduction in the single quantum limit [6]. Coherent coupling of gigahertz (GHz) acoustic phonons with both GHz and optical photons has resulted in a coherent microwave-to-optical transduction at cryogenic temperature [7–12]. Here, we predict a coherent coupling of GHz acoustic phonons with a type of quasiparticles called ferrons [13–15], along with several new physical phenomena enabled by such new coupling.

Ferrons were introduced theoretically [13–15] as the elementary excitation of electric polarization in ferroelectrics, by analogy to magnons (the elementary excitation of spins) [16]. Incoherent ferrons represent the collective amplitude of the polarization fluctuation and do not collectively have well-defined frequency and phase [17]. Coherent ferrons refer to the quanta of polarization waves [18–21] that collectively oscillate at a single frequency with well-defined phase. The fundamental-mode (i.e., the wavenumber is zero) coherent ferrons (akin to the fundamental-mode magnons [3,22–27]) represent the coherent and in-phase oscillation of electric dipoles that are spatially uniform at the ground state. In many ferroelectrics, the fundamental-mode coherent ferron is also a ferroelectric soft mode [15,20], which is the lowest-frequency polar optical phonon mode in the ferroelectric phase whose resonant frequency decreases substantially (i.e., the force constant becomes softer) as temperature increases to be near the ferroelectric-to-paraelectric phase transition [28,29].

There are three main advantages in exploring the fundamental-mode coherent ferrons for hybrid quantum systems. First, since the resonant ferron-photon coupling is based on electric dipole interactions, the coupling strength can be several orders of magnitude stronger than hybrid systems based on magnetic dipole interactions [14,20], e.g., a hybrid magnon-photonic system. Second, the resonant frequency ($\omega_0$) of coherent ferrons can reach tens of GHz to terahertz (THz) regime [30–32] without the need of any strong bias electric fields. Such high frequencies lead to a reduced occupation number ($\bar{n}\approx k_B T/(\hbar\omega_0)$ compared to a few GHz excitations, where $k_B$ is the Boltzmann constant, $\hbar$ is the reduced Planck constant) at a given temperature $T$, thereby easing the refrigeration requirement for reaching the quantum ground state ($\bar{n}\ll 1$) [33]. Third, the polarization nature of ferrons allows controlling ferron-based coherent states using an electric field, which is easy to localize on a chip as opposed to a magnetic field.

Cavity acoustic phonons have recently emerged as highly promising building blocks for quantum hardware [34–37] thanks to their coherent coupling to superconducting qubits [35–40] and their significantly smaller wavelength than free-space photons. Achieving a strong coupling between the fundamental-mode coherent ferrons and cavity acoustic phonons will potentially enable a hybrid quantum system that combines the unique advantages of both quasiparticles for realizing new control modalities.

In this Letter, we theoretically demonstrated this highly desirable state with strongly coupled ferrons and bulk cavity acoustic phonons in a nanometer (nm)-thick freestanding ferroelectric membrane, which concurrently functions as a cavity for both the ferrons and acoustic phonons.

Using a van der Waals ferroelectric $CuInP_2S_6$ (CIPS) membrane as an example, we predict a strong to ultra-strong coupling between the ferrons and cavity bulk acoustic phonons, as well as the capability to tune the coupling by temperature, electric field, and strain, and notably, a new control modality originating from ferroelectric switching. Furthermore, near the ferroelectric-to-paraelectric phase transition of CIPS, we show that an applied strain can drive the hybrid ferron-phonon system into the deep-strong coupling regime, with $g_c/\omega_0>1$, where $\omega_0$ is the resonant frequency of the coherent ferrons and cavity acoustic phonons.

***Ferron Excitation***. The ferrielectric CIPS can be considered as a uniaxial ferroelectric with a polar axis along the $x_3$ axis [17], similarly to canonical uniaxial ferroelectric materials such as $LiNbO_3$ and AlScN. Figure 1(a) schematically shows the out-of-plane translation of $Cu^+$ and $In^{3+}$ cations (replotted based on the lattice parameters and atomic coordinates from first-principles calculations [41]), yielding a net out-of-plane polarization. Here, we consider CIPS membrane as an example for three reasons. First, the resonant frequency of the fundamental-mode coherent ferron in the CIPS, which represents the dynamics of a collective polarization coordinate (see details in Sec. 1 in Supplemental Materials [42] and references therein [17,41,43–53]), is tens of GHz at 293 K [17,45]. This resonant frequency is close to the frequencies of low-order bulk cavity acoustic phonons in a nm-thick CIPS membrane at room temperature, making it possible to achieve strong ferron-phonon coupling at room temperature. By comparison, for $LiNbO_3$ and AlScN, the resonant frequencies of $A_1$ transverse optical phonon (related to their spontaneous polarization) are approximately 7.3 THz [31,54] and 18.3 THz [46,55], respectively, at 298 K, far exceeding the upper-bound frequencies of acoustic phonons in these materials. Second, CIPS has large electrostrictive coefficients (e.g., $Q_{33}\approx$ -1.79 $m^4/C^2$ for CIPS [42], as compared to $Q_{33}\approx$ 0.13 $m^4/C^2$ for $Al_{0.5}Sc_{0.5}N$ and $Q_{33}\approx$ -0.01 $m^4/C^2$ for $LiNbO_3$ [46]) and a robust equilibrium polarization [17]. These two properties result in a large static piezoelectric coupling coefficient (which is a linear function of electrostrictive coefficient and polarization [56]). This will in turn lead to a large ferron-phonon coupling strength, which is defined as the rate of energy exchange between the ferrons and acoustic phonons and has units of Hz. As will be derived later in this paper, the ferron-phonon coupling strength is linearly proportional to the piezoelectric/electrostrictive coefficient. Third, nm-thick CIPS membranes can be conveniently obtained via mechanical exfoliation from bulk single crystals [47,57].

We further consider the excitation of the fundamental-mode coherent ferrons in a freestanding nm-thick CIPS membrane by a microwave field and study its coupling to bulk cavity acoustic phonons. The electric-field component of the incident microwave field is $E_i^{\mathrm{inc}}=E_i^0e^{-\mathrm{i}\omega t}$, where $E_i^0$ is the real-valued field amplitude. We assume that $E_i^{\mathrm{inc}}$ is spatially uniform inside the thin CIPS membrane, which should facilitate the direct excitation of the fundamental-mode coherent ferrons, although higher-order ($k=n\pi/d$, $n$=1, 2, 3, .., and $d$ is the thickness) coherent ferrons would be excited indirectly by cavity acoustic phonons via piezoelectric coupling, as illustrated in Fig. 1(b).

The ferron dispersion relation takes the form of $\omega^2=\omega_{\mathrm{f}}^2+(G_0/\mu)k^2$ [15,58], which can be readily derived by linearizing the equation of motion for lattice polarization (see Sec. 2 in Supplemental Materials [42] and references therein [32,59,60]). Here, $\omega_{\mathrm{f}}$ is the resonant frequency of the fundamental-mode ferron, $G_0=G_{3333}$ is the relevant component of the gradient energy coefficient tensor (see Sec. 1 in Supplemental Material [42]), and $\mu$ is the effective mass coefficient. In archetypal ferroelectrics, the gradient energy coefficients can be determined based on the dispersion relations of the soft mode phonons in the paraelectric phase [49]). In the ferrielectric

CIPS, it is less straightforward to determine gradient energy coefficients via this approach, because the spontaneous polarization is a collective polarization coordinate. To our knowledge, the value of $G_0$ for CIPS has not yet been rigorously determined. Figure 1(c) shows the analytically calculated $\omega$-$k$ relation of ferrons by increasing $G_0$ over a wide range from $5\times10^{-10}$ to $5\times10^{-6}$ $m^3/C^2$. The $\omega$-$k$ relation of the longitudinal acoustic (LA) phonon, given by $\omega = v_{LA}k$, is also plotted in Fig. 1(c). Here, the velocity of LA phonons is approximately calculated as $v_{LA} = \sqrt{c_{33}/\rho}$, where $c_{33}$ is the bare (i.e., determined under zero polarization) elastic stiffness coefficient and $\rho$ is the mass density. As shown in Fig. 1(c), the frequency of the $n$=1 mode ferron is only slightly higher than the fundamental-mode ferron when $G_0$ is small, displaying a flat-band-like feature. In this case, it is possible to realize the tripartite coupling among the fundamental-mode ferron, $n$=1 mode ferron, and $n$=1 mode phonons. When $G_0$ is large, the frequency of the $n$=1 mode ferron is significantly higher than both the fundamental-mode ferron and the $n$=1 mode phonon. Moreover, a large $G_0$ favors a spatially uniform polarization, yielding a small polarization amplitude for the $n$=1 mode ferron. Thus, the strong coupling would primarily occur between the fundamental-mode ferron and the $n$=1 mode phonon. For a fixed membrane thickness, one can also tune $\omega_f$ by varying the temperature, bias electric field, and strain, such that $\omega_f$ can match the frequency of the higher-order ($n$=2, 3,..) acoustic phonon modes. The focus of this work is on studying the strong coupling between the fundamental-mode ferron and these cavity acoustic phonons ($n$=1,2,3,..) by considering a relatively large $G_0$ to suppress the $k\neq0$ coherent ferron modes. This will lay the theoretical groundwork for investigating and harnessing ferron-phonon coupling and other ferron-based hybrid states in the finite-$k$ regime.

At the ferron resonance ($\omega=\omega_f$), $E_i^{\mathrm{inc}}$ will be absorbed strongly. The hybridization of ferrons and bulk acoustic phonons creates a nonzero frequency gap (mode split) of $\omega_+ - \omega_-$, and the absorption of $E_i^{\mathrm{inc}}$ occurs at $\omega=\omega_\pm$. Therefore, frequency-dependent power absorption spectrum of the microwave field, $P_{\mathrm{abs}}(\omega)$, can be used to quantify the frequency gap (which is related to the ferron-phonon coupling strength $g_c$) and the dissipation rates $\kappa_f$ and $\kappa_{\mathrm{ph}}$. Theoretically, the power absorption spectrum can be evaluated as $P_{\mathrm{abs}} \propto \mathrm{Im}\left(E_i^{\mathrm{inc},*}\Delta P_i\right) = \mathrm{Im}\left(E_i^0 \chi_{ij} E_j^0\right)$, where $E_i^{\mathrm{inc},*} = E_i^0 e^{\mathbf{i}\omega t}$ is the complex conjugate of $E_i^{\mathrm{inc}}$, 'Im' denotes the imaginary component, $\chi_{ij}$ is the linear susceptibility, with $i, j$ = 1, 2, 3 indicating the three orthogonal axes in the crystal physics coordinate system of the CIPS. The electric field-induced lattice polarization is given by $\Delta P_i = \chi_{ij} E_j^{\mathrm{inc}}$, with $\Delta P_i = P_i - P_i^{\mathrm{eq}}$, and $P_i^{\mathrm{eq}}$ is the lattice polarization at thermodynamic equilibrium. If $E_i^{\mathrm{inc}}$ only contains a $z$-component ($z\|x_3$), $P_{\mathrm{abs}} \propto \mathrm{Im}(\chi_{33})$.

The analytical expression of $\chi_{33}(\omega)$ can be derived by linearizing the coupled equations of motion for lattice polarization and mechanical displacement under a traction-free boundary condition at the top and bottom surfaces of the membrane, given by (see details in Sec. 2 in Supplemental Material [42] and references therein [32,59,60]).

$$\chi_{33}(\omega) \approx \frac{1}{\kappa_0} \frac{1}{\mu(\omega_f^2 - \omega^2) - \mathbf{i}\gamma\omega + L_{333}\Omega_{333}}, \tag{1}$$

where $\kappa_0$ is the vacuum permittivity, $\omega_f = \sqrt{K_{33}/\mu}$, $K_{33} = \partial^2 f^{\mathrm{E}}/\partial P_3^2$ is the local curvature of the free energy landscape at $P_3 = P_3^{\mathrm{eq}}$ [31], where $f^{\mathrm{E}}$ is the Helmholtz free energy density (see Sec. 1 in Supplemental Material [42]). $\gamma$ is the polarization damping coefficient. $L_{333} = \partial^2 f^{\mathrm{E}}/(\partial P_3 \partial \varepsilon_{33})$ describes the piezoelectric coupling between $P_3$ and the total strain $\varepsilon_{33}$ at $P_3 = P_3^{\mathrm{eq}}$. $\Omega_{333}(\omega)$=$\langle\Delta\varepsilon_{33}\rangle/\Delta P_3$ is defined as the electromechanical susceptibility, where $\langle\Delta\varepsilon_{33}\rangle$ is the spatial average of $\Delta\varepsilon_{33}(x_3)$. For CIPS, one has $L_{333} \approx -2c_{33}Q_{33}P_3^{\mathrm{eq}}$, $\Omega_{333}$=$-\frac{4L_{333}}{A_0\rho v_{\mathrm{LA}}}\tanh\left(\frac{A_0}{4v_{\mathrm{LA}}}\right)$, where $A_0$=$d\omega(\beta\omega - 2\mathbf{i})$ and $\beta$ is the elastic damping coefficient. The velocities of longitudinal and transverse acoustic phonons are $v_{\mathrm{LA}}$=$\sqrt{c_{33}/\rho}$ and $v_{\mathrm{TA}}$=$\sqrt{c_{55}/\rho}$, respectively. $c_{33}$, $c_{35}$, and $c_{55}$ are the bare elastic stiffness coefficients, $Q_{33}$ is the electrostrictive coefficient.

We note that Eq. (1) is only applicable to a system that is dominated by the coupling between the fundamental-mode ferrons and cavity acoustic phonons. If we consider the coupling between $k\neq 0$ coherent ferrons and acoustic phonons, the elastic stiffness tensor would be wavenumber-dependent, complex-valued, and determined by the value of lattice polarization and the strength of dynamical piezoelectric coupling. In that regard, the resulting sound velocities and resonant frequencies of acoustic phonons would vary with the temperature, electric field, and strain. Detailed derivation and discussion are provided in Secs. 1,3,4 in Supplemental Material [42] with supporting references [17,41,43–53,61].

As further shown in Fig. 1(d), the local curvature of the free energy landscape decreases as temperature ($T$) increases, yielding a reduced $\omega_{\mathrm{f}}$. Near the ferrielectric-to-paraelectric phase transition temperature of ~ 312 K, $\omega_{\mathrm{f}}$ approaches zero due to the almost zero curvature, as shown in Fig. 1(e). $P_3^{\mathrm{eq}}$ decreases concomitantly with increasing temperature before dropping abruptly to zero at 312 K and above, as indicated by the shifting energy minima.

***Ferron-phonon coupling* strength at resonance**. The frequencies of longitudinal cavity bulk acoustic phonons are $\omega_n^{\mathrm{ph}} = \frac{n\pi}{d} v_{\mathrm{LA}}$. The fundamental-mode ferrons can only have non-zero coupling with odd-numbered cavity acoustic phonons. At resonance, i.e., $\omega_{\mathrm{f}}$=$\omega_n^{\mathrm{ph}}$=$\omega_0$, the ferron-phonon coupling strength $g_c$ can be derived based on Eq. (1),

$$g_c = \frac{\sqrt{2}|L_{333}|}{d\omega_0\sqrt{\rho\mu}} \approx \frac{2\left|Q_{33}P_3^{\mathrm{eq}}\right|}{n\pi}\sqrt{\frac{2c_{33}}{\mu}}, \tag{2}$$

where $n$ = 1, 3, 5… is an odd integer number. Detailed derivation of Eq. (2) is given in Sec. 4 of Supplemental Material [42] with supporting references [59,61]. Thus, $g_c$ should increase linearly with $\left|P_3^{\mathrm{eq}}\right|$, which characterizes the volumetric density of electric dipoles. This contrasts with the magnon-based hybrid systems where $g_c \propto \sqrt{M_s}$ [62], with $M_s$ (saturation magnetization) characterizing the volumetric spin density. Equation (2) also indicates that $g_c$ is linearly proportional to the term $2\left|Q_{33}P_3^{\mathrm{eq}}\right|$, which approximately represents the magnitude of piezoelectric voltage coefficient $d_{33}^{\sigma} = (\partial E_3/\partial\sigma_{33})_D$, where $\sigma_{33}$ is stress and $D$ is the electric displacement. Furthermore, $g_c$ is inversely proportional to the order of acoustic phonons ($n$) due to the greater overlap in the spatial profiles of ferron and phonons at lower $n$ values. At resonance, $\chi_{33}$ develops two pairs of conjugated poles, $\omega_{\pm}$, at which Im($\chi_{33}$) is maximized, and frequency gap between these two peaks is $\omega_+ - \omega_-$. In the strong coupling regime, where $g_c/\kappa_{\mathrm{f}}$>1 and $g_c/\kappa_{\mathrm{ph}}$>1 but $g_c/\omega_0$<0.1 [59], one has $g_c$≈$(\omega_+ - \omega_-)$/2.

As an example, Figure 1(f) shows the frequency- and temperature- dependent Im($\chi_{33}$) in a 48.9 nm CIPS film calculated based on Eq. (1), with $\gamma$=$10^{-3}$ Ω·m (this value was reported in [17] and determined by fitting the experimentally-measured temperature dependence of polarization relaxation time near the ferrielectric-to-paraelectric phase transition [63]) and $\beta$ = 9.19×10$^{-14}$ s (which is extracted based on the linewidth of the longitudinal acoustic phonon resonance measured by Brillouin light scattering [64], see Sec. 5 in Supplemental Materials [42] and references therein [64,65]). Within 0-310 K, the ferron resonance frequency $\omega_\mathrm{f}/2\pi$ varies from 135.7 GHz to 30.0 GHz. At 298 K, $\omega_\mathrm{f}/2\pi$=$\omega_{n=1}^{\mathrm{ph}}/2\pi$=46.0 GHz, and a large $g_c/2\pi$ of 7.27 GHz is calculated via Eq. (2), which is close to the half of the frequency gap of 7.50 GHz extracted from Fig. 1(e).

The dissipation rates of uncoupled ferron and phonons, $\kappa_\mathrm{f}$ and $\kappa_\mathrm{ph}$, are defined as the half-width-half-maximum linewidths of the power absorption spectrum in a pure ferron system and the phononic branch of the spectrum in a hybrid ferron-phonon system, respectively. Based on Fig. 1(f), one has $\kappa_\mathrm{f}/2\pi$=1.05 GHz and $\kappa_\mathrm{ph}/2\pi$ =0.65 GHz. The dissipation rates can also be estimated analytically via $\kappa_\mathrm{f} \approx \frac{\gamma}{2\mu}$ and $\kappa_\mathrm{ph} \approx \frac{\beta\omega_0^2}{2}$ (see Sec. 4 in Supplemental Materials [42]), resulting in a $\kappa_\mathrm{f}/2\pi$ of 1.00 GHz and a $\kappa_\mathrm{ph}/2\pi$ of 0.61 GHz that are consistent with the extracted values. Using $g_c/2\pi$=7.27 GHz, and the extracted $\kappa_\mathrm{f}$ and $\kappa_\mathrm{ph}$, we obtain a cooperativity $C$=$g_c^2/(\kappa_\mathrm{f}\kappa_\mathrm{ph})$ of 86.64, which is comparable to the cooperativities reported in hybrid magnon-phonon systems [66–71]. Notably, since $g_c$ exceeds both $\kappa_\mathrm{f}$ and $\kappa_\mathrm{ph}$, the hybrid ferron-phonon system is in the strong coupling regime [5]. Furthermore, since the system has a $g_\mathrm{c}/\omega_0$ = 0.16>0.1, the system falls in the ultra-strong (USC) coupling regime according to [59]. By comparison, the $g_\mathrm{c}/\omega_0$ ratios for typical hybrid magnon-phonon systems remain well below 0.1. This is partly because spontaneous electrostriction (~10$^{-2}$), which characterize the polarization-strain coupling, is generally larger than the spontaneous magnetostriction (10$^{-5}$~10$^{-3}$) that describes the magnetization-strain coupling [72].

Figure 1(e) further shows the evolution of the changes in the intrinsic energy densities of the ferron and phonon systems, upon the excitation of the CIPS membrane by a Gaussian-enveloped electric field pulse $E_3^{\mathrm{inc}}(t)$ with a center frequency of 46.0 GHz at $t$=0. The intrinsic energy density of the ferron system contains the kinetic energy density of ferrons and the free energy density terms that do not involve direct coupling to strain, and likewise for the phonon system. Details of time-domain solution and energy analyses are provided in Sec. 6-7 of Supplemental Materials [42] and references therein [30–32,46]. A complete energy transduction, i.e., the maximum energy change in one system corresponds to zero change in the other, is shown in Fig. 1(g). Such a Rabi-like oscillation process is a typical time-domain feature for the strong coupling [5,73]. The period of such oscillation is approximately 69.0 ps, corresponding a frequency of ~14.5 GHz, which is equal to the frequency gap of 15.0 GHz extracted from Fig. 1(e) at $\omega/2\pi$=46.0 GHz.

***Electric-field control of ferron-phonon coupling***. Under a fixed membrane thickness $d$, the resonant frequencies of cavity bulk acoustic phonons $\omega_n^{\mathrm{ph}}$ are fixed. Applying a bias electric field along the thickness direction $E_3$, which can be achieved without electrodes (as in [33]), can simultaneously tune the local curvature (i.e., $K_{33}$) at $P_3$=$P_3^{\mathrm{eq}}$ and the value of $P_3^{\mathrm{eq}}$. When $E_3$ exceeds the coercive electric field, the polarity of $P_3^{\mathrm{eq}}$ is reversed. This process can be seen from the $E_3$-dependent free energy profiles in Fig. 2(a), and the $P_3^{\mathrm{eq}} - E_3$ hysteresis loop shown in Fig. 2(b). Thus, a bias electric field can modulate the ferron-phonon coupling by detuning the $\omega_\mathrm{f}$ from the resonance condition ($\omega_\mathrm{f}$=$\omega_n^{\mathrm{ph}}$=$\omega_0$) and modulating $|P_3^{\mathrm{eq}}|$. Figure 2(c) shows the power absorption spectrum of a 48.9-nm-thick CIPS membrane calculated via Eq. (1) by sweeping $E_3$ from -0.1 MV/cm to 0.1 MV/cm at 298 K. At $E_3$=0 MV/cm, $\omega_\mathrm{f}/2\pi$ is still 46.0 GHz, which is equal to $\omega_{n=1}^{\mathrm{ph}}$ ($\equiv \omega_1$). When $E_3$ approaches the coercive field (0.068 MV/cm), $\omega_{\mathrm{f}}/2\pi$ decreases rapidly to 14 GHz due to the flattened energy landscape [see Fig. 2(a)]. As $E_3$ exceeds the coercive field, $P_3^{\mathrm{eq}}$ falls to the other energy minimum with a sudden increase in the local curvature of the energy profile, leading to a jump of $\omega_{\mathrm{f}}/2\pi$ to 56.3 GHz. The electric-field control of $\omega_{\mathrm{f}}$ is shown by the dashed curves in Fig. 2(c). As $\omega_{\mathrm{f}}$ is tuned further away from $\omega_1$, the absorption spectrum becomes closer to those of the uncoupled ferrons and acoustic phonons. This trend is quantitatively shown by the electric field-dependent frequency offset with respect to the uncoupled ferron, $\Delta\omega_{\mathrm{f}}$, and $n$=1 mode cavity acoustic phonon, $\Delta\omega_1$. As shown in Fig. 2(d,e), both the $\Delta\omega_{\mathrm{f}} - E_3$ and $\Delta\omega_1 - E_3$ curves display hysteric behaviors due to the $P_3^{\mathrm{eq}} - E_3$ hysteresis. If $E_3$ is kept below the coercive electric field, reversible changes in $P_3^{\mathrm{eq}}$, $\Delta\omega_{\mathrm{f}}$, and $\Delta\omega_1$ are obtained, as shown by the red curves in Figs. 2(b,d,e), respectively. $\Delta\omega_{\mathrm{f}}$ reaches its maximum at $E_3$=0, where $\omega_{\mathrm{f}}$=$\omega_1$. The discontinuity in both the $\Delta\omega_{\mathrm{f}}$ and $\Delta\omega_1$ at $E_3$=0 arises from the asymmetric frequency gap in an USC regime, i.e., $|\omega_+ - \omega_0| \neq |\omega_- - \omega_0|$ on resonance. These results emphasize the capability of using an electric field to activate/deactivate the ferron-phonon hybridization or achieve distinct levels of coupling under an identical electric field. Such bistable control, arising from the ferroelectric switching behavior, represents a new control modality for a hybrid quantum system that can be challenging to achieve through existing technologies.

An electric field can also be used to enable mode-specific ferron-phonon hybridization with various level of coupling strength $g_c$ in thicker CIPS membranes. For example, Figure 3(a) shows the power absorption spectrum of a 120-nm-thick CIPS membrane by sweeping the applied bias electric field $E_3$ from -0.1 MV/cm to 0.1 MV/cm at 298 K. Figure 3(b) shows the corresponding frequency offset with respect to the uncoupled ferron, $\Delta\omega_{\mathrm{f}}$, and to $n$=1,3 mode cavity acoustic phonons, $\Delta\omega_{1,3}$. Specifically, near $E_3$=-0.066 MV/cm, where $\omega_{\mathrm{f}}/2\pi$=$\omega_3/2\pi$=$\omega_0/2\pi$=56.0 GHz, ferrons are strongly coupled to $n$=3 mode phonon, with a coupling strength $g_{\mathrm{c}}/2\pi$ =2.66 GHz. The hybrid system is in the strong coupling (SC) regime, because $g_{\mathrm{c}}$ is greater than both $\kappa_{\mathrm{f}}$ and $\kappa_{\mathrm{ph}}$, and because $0<g_{\mathrm{c}}/\omega_0<0.1$. When $E_3$ approaches the coercive electric field at 0.066 MV/cm, where $\omega_{\mathrm{f}}/2\pi$=$\omega_1/2\pi$=$\omega_0/2\pi$=18.7 GHz, ferrons are strongly coupled to the $n$=1 mode phonon, with $g_{\mathrm{c}}/2\pi$=7.33 GHz, $\kappa_{\mathrm{f}}/2\pi$=0.99 GHz, and $\kappa_{\mathrm{ph}}/2\pi$=0.10 GHz. The system is in the USC regime since $0.1<g_{\mathrm{c}}/\omega_0<1$. When $E_3$ exceeds the coercive electric field, the reversal of $P_3^{\mathrm{eq}}$ leads to the sudden increase of $\omega_{\mathrm{f}}$ to 56 GHz, thus turning the hybrid system into the SC regime. Figure 3(c-d) provides another example in an even thicker CIPS membrane ($d$=200 nm), where varying $E_3$ can selectively activate the strong or ultra-strong coupling between fundamental-mode ferrons and $n$=1,3,5 mode phonons, and lead to rapid switching from $n$=1 mode to $n$=5 mode coupling at near the coercive electric field. The ferroelectric switching therefore provides a rich spectrum of new functionalities for the control of hybrid quantum systems.

**Strain-enabled multimode ferron-phonon deep-strong coupling**. In addition to electric field, strain can also modulate $\omega_{\mathrm{f}}$ and hence the ferron-phonon coupling. One notable difference is that strain can induce a ferrielectric-to-paraelectric phase transition, where $P_3^{\mathrm{eq}}$ suddenly decreases to zero when strain exceeds a threshold. This is shown by the free energy profiles under different strains, $\varepsilon^{\mathrm{app}}$, applied along the $x_1$ axis in Fig. 4(a). The treatment on the mechanical boundary condition of a uniaxially strained ferroelectric nanomembrane is discussed in [46]. Different from room-temperature (298 K) operation in Figs. 2 and 3, here we set the temperature to 305 K (i.e., closer to the ferrielectric-to-paraelectric transition temperature of 312 K in a mechanically free CIPS membrane), because the effect of strain on $\omega_{\mathrm{f}}$ is more pronounced near the phase transition.

Importantly, we show that the applied strain can enable a fundamentally intriguing multimode deep-strong coupling (DSC) between ferrons and cavity bulk acoustic phonons, with $g_c/\omega_0$>1, i.e., the rate of energy exchange between the two systems is faster than the eigenfrequencies of uncoupled modes. As one example, Figure 4(b) shows the strain- and frequency-dependent power absorption spectrum of a 500-nm-thick CIPS membrane at 305 K, and the corresponding strain-dependent frequency offsets with respect to the uncoupled modes are shown in Fig. 4(c). When 2.3%< $\varepsilon^{\mathrm{app}}$ <2.7%, $\Delta\omega_{\mathrm{f}}$, $\Delta\omega_1$, and $\Delta\omega_3$ are simultaneously large, indicating a multimode ferron-phonon coupling. Notably, at $\varepsilon^{\mathrm{app}}$=2.65%, where $\omega_{\mathrm{f}}/2\pi$ =$\omega_1/2\pi$=$\omega_0/2\pi$=4.37 GHz, the lower branch of the absorption spectrum disappears, which is a hallmark feature of DSC as has been reported in a hybrid electron–photon system [74]. The coupling strength $g_c/2\pi$ calculated via Eq. (2) is 5.60 GHz, resulting in a $g_c/\omega_0$=1.64>1, confirming the DSC condition.

***Conclusion***. We have theoretically demonstrated a tunable coherent coupling between the fundamental-mode ferrons and cavity bulk acoustic phonons in a freestanding ferroelectric membrane, using the $CuInP_2S_6$ as an example. We present analytical formulae that connect experimentally measurable material parameters to the coupling strength $g_c$ and dissipation rates ($\kappa_{\mathrm{f}}$ and $\kappa_{\mathrm{ph}}$). These analytical formulae highlight the fundamental difference between such ferron-phonon coupling and its counterpart of magnon-phonon coupling [66–71]. From an application perspective, our findings demonstrate the prospects of utilizing coherent ferrons for electric field-controllable hybrid quantum systems that reach the ultra-strong and even deep-strong coupling regime. Such a hybrid ferron-phonon system enables hysteric and bistable ferroelectric switching, thereby providing new possibilities for quantum transduction, computing, and sensing. Overall, we believe our findings should be of broad interests to readers in the field of ferroelectrics, piezoelectrics, magnonics, and hybrid dynamical/quantum systems.

Although our present theoretical calculations are based on ferroelectric systems hosting primarily fundamental-mode ($k$=0) coherent ferrons, the analytical model can be generalized to understand and predict the ferron-phonon coupling in the more complex finite-$k$ regime. One example is the strong tripartite coupling among $k$=0 coherent ferrons, $k=\pi/d$ mode coherent ferrons, and $k=\pi/d$ mode acoustic phonons, which can occur when the gradient energy coefficient is relatively small based on our dynamical phase-field simulations (see Sec. 8 in Supplemental Materials [42]). From the perspective of experimental design, our calculation results are applicable to ferroelectric membranes on top of an elastically compliant substrate (e.g., polymer) which can also reflect acoustic waves for creating cavity bulk acoustic phonons in the ferroelectric. Our analytical model can also be extended to other experimentally feasible structures such as a bilayer ferroelectric/SiN membrane on a polymer substrate where SiN can provide additional mechanical support, or a freestanding membrane on an elastically stiff substrate (e.g., Si) patterned with air cavity [75]. Our numerical model (i.e., the dynamical phase-field model, see brief introduction in Sec. 6 in Supplemental Materials [42]) can be utilized to study the ferron-phonon coupling in three dimensions such as the possible strong coupling between coherent ferrons and surface acoustic waves (SAW, analogous to magnon-SAW hybridization [76]), as well as the interaction between cavity acoustic phonons and the collective modes of ferroelectric domain walls [63,64], polar vortices [77], polar skyrmions [78,79], and other topologically nontrivial polar textures in other ferroelectric and polar materials within the framework of Landau-Ginzburg-Devonshire theory.

**Acknowledgements**

This work is primarily supported by the US Department of Energy, Office of Science, Basic Energy Sciences, under Award Number DE-SC0020145 as part of the Computational Materials Sciences Program (Y.Z., A.N.M., E.A.E., V.G., L.-Q.C., and J.-M.H.). X. Z. acknowledges support from National Science Foundation (NSF) under Grant No. ECCS-2337713. W. Z. and J.-M.H. also acknowledge support from the NSF under Grant No. DMR-2509513. Dynamical phase-field simulations in this work are supported by the NSF under Grant No. DMR-2237884 and were performed using Bridges at the Pittsburgh Supercomputing Center through allocation TG-DMR180076 from the Advanced Cyberinfrastructure Coordination Ecosystem: Services and Support (ACCESS) program, which is supported by NSF Grants No. 2138259, No. 2138286, No. 2138307, No. 2137603, and No. 2138296. Partial support for manuscript preparation was provided by the Wisconsin MRSEC (DMR-230900). Partial support of A.N.M. and E.A.E. work was provided by the National Academy of sciences of Ukraine.

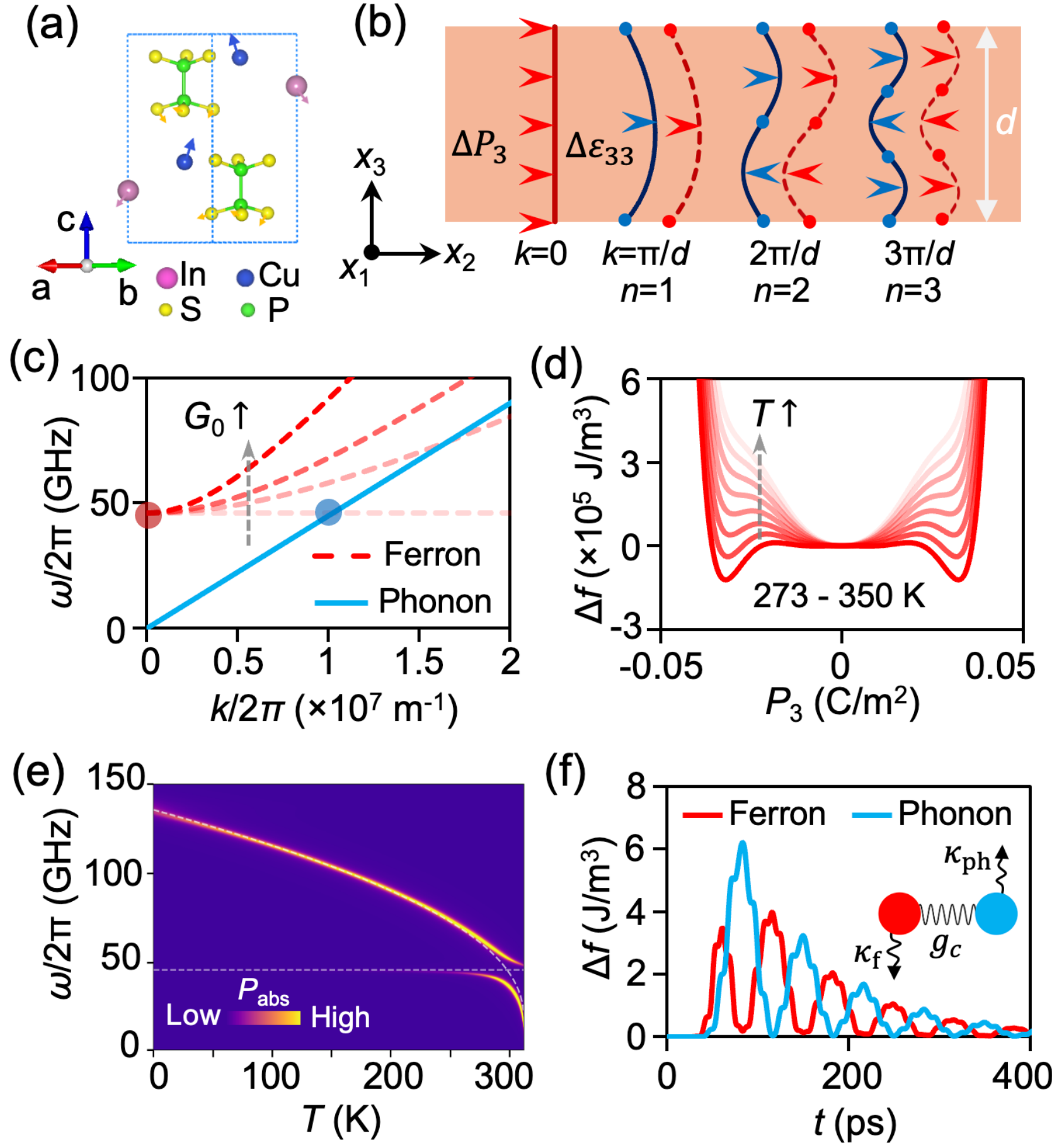

**Figure 1.** (**a**) Unit cell of $CuInP_2S_6$ (CIPS), where the out-of-plane translations of $Cu^+$ and $In^{3+}$ atoms yield a net spontaneous polarization along the $c$ ($x_3$) axis. The arrows indicate the atomic displacement vectors corresponding to the ferroelectric soft mode. (**b**) Spatial profiles of the fundamental-mode ($k$=0) and higher-order ($k=n\pi/d$, $n$=1,2,3…) coherent ferrons and cavity bulk acoustic phonons in a freestanding CIPS membrane with a thickness $d$. (**c**) Analytically calculated dispersion relations of the uncoupled coherent ferrons and longitudinal acoustic phonons for CIPS with $G_0$=5×10$^{-10}$, 1×10$^{-6}$, 2×10$^{-6}$ , and 5×10$^{-6}$ J m$^3$/C$^2$. The red and blue dots denote the fundamental-mode ($k$=0) ferron and the $n$=1 mode ($k=\pi/d$) cavity bulk acoustic phonon mode in a CIPS membrane ($d$=48.9 nm), respectively. (**d**) Temperature-dependent free energy density $\Delta f$ of a stress-free CIPS membrane. (**e**) Temperature- and frequency- dependent power absorption spectrum of the incident microwave field, $P_{abs}$, in a 48.9-nm-thick CIPS membrane. The dashed lines indicate the temperature-dependent resonant frequencies of the uncoupled ferrons and $n$=1 mode bulk acoustic phonons. (**f**) Evolution of the intrinsic energy densities of the ferron and phonon systems excited by a Gaussian-enveloped microwave pulse at 298 K. The inset illustrates the hybrid ferron-phonon system with a coupling strength $g_c$ and dissipation rates $\kappa_f$ and $\kappa_{ph}$.

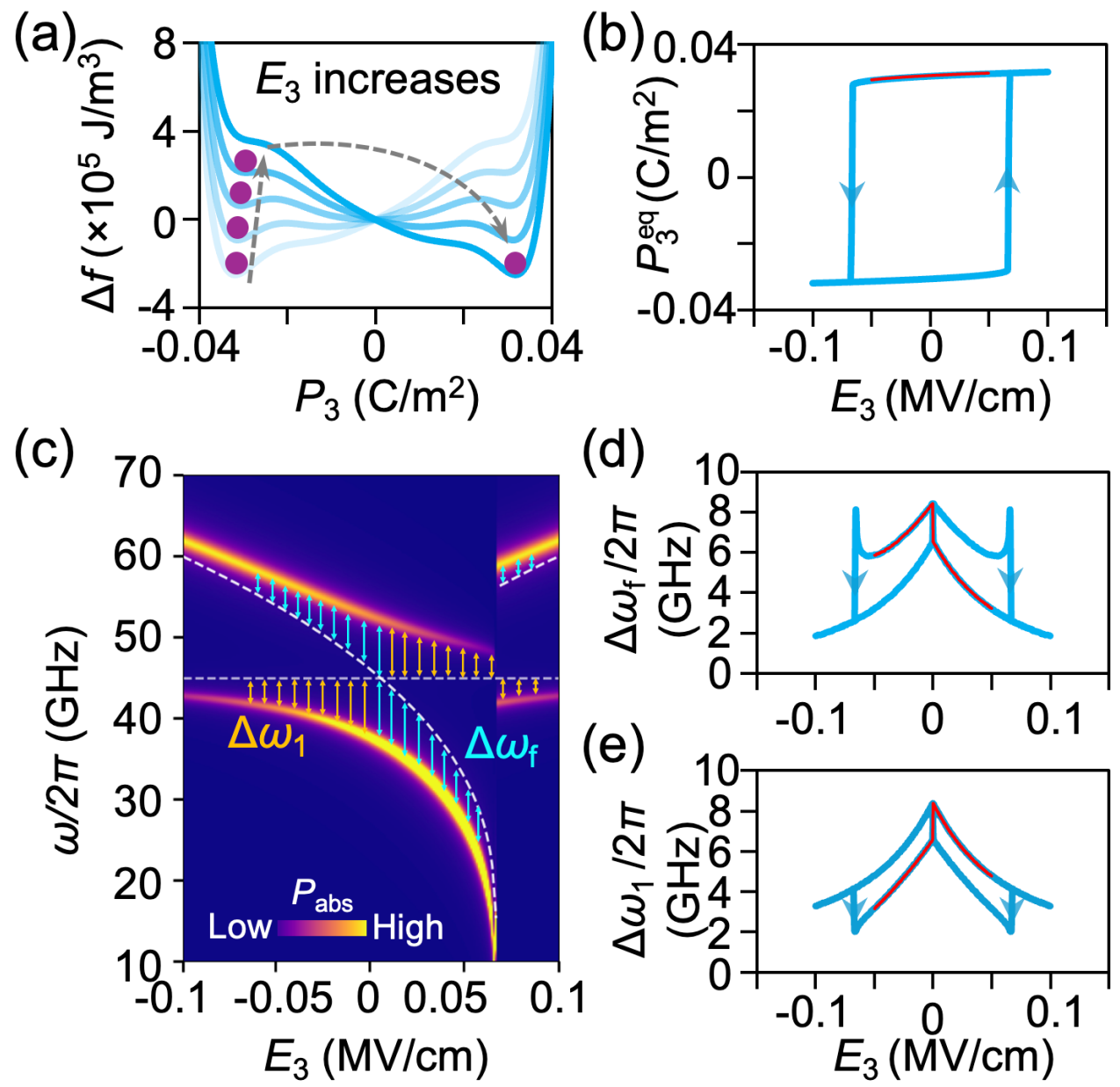

**Figure 2**. (**a**) Electrical field ($E_3$) dependent 1D free energy density $\Delta f$ in a mechanically free CIPS membrane, where $E_3$ increases from -0.1 MV/cm to 0.1 MV/cm at 298 K. Purple dots indicate local energy minima. (**b**) $P_3^{\mathrm{eq}} - E_3$ hysteresis loop calculated by sweeping $E_3$ from -0.1 MV/cm to 0.1 MV/cm and then back to -0.1 MV/cm at 298 K. The red line ($|E_3| \leqslant 5$ MV/m) indicates a reversible polarization switching with an initial polarization along $+x_3$ ($P_3^{\mathrm{eq}}>0$). (**c**) $E_3$- and frequency-dependent power absorption spectrum of the driving microwave field, $P_{\mathrm{abs}}$, in a 48.9-nm-thick CIPS membrane, where $E_3$ increases from -0.1 MV/cm to 0.1 MV/cm at 298 K. The dashed curve and horizontal line indicate the $E_3$-dependent resonant frequencies of the uncoupled coherent ferrons, $\omega_{\mathrm{f}}$, and $n$=1 mode bulk acoustic phonons, $\omega_1$, respectively. The blue and orange arrows indicate the frequency offset with respect to the uncoupled ferron ($\Delta\omega_{\mathrm{f}}$) and $n$=1 mode phonon ($\Delta\omega_1$), respectively. (**d, e**) $E_3$-dependent $\Delta\omega_{\mathrm{f}}$ and $\Delta\omega_1$, corresponding to the electric field sweeping sequence in (**b**).

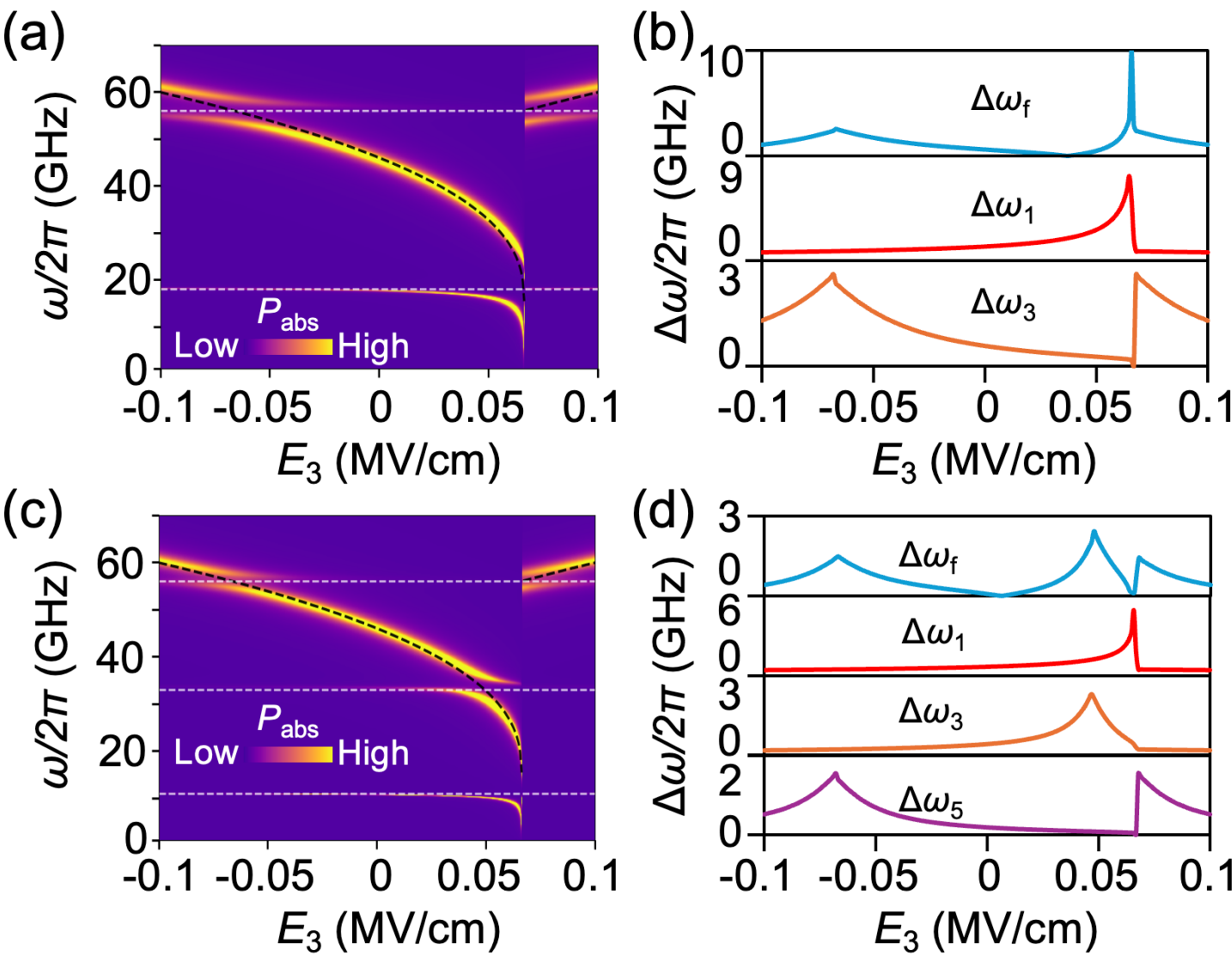

**Figure 3**. $E_3$- and frequency-dependent power absorption spectra of the driving microwave field, $P_{abs}$, in a (**a**) 120-nm-thick and (**c**) 200-nm-thick mechanically free CIPS membrane, where $E_3$ increases from -0.1 MV/cm to 0.1 MV/cm at 298 K. The black dashed curve indicates the $E_3$-dependent resonant frequencies of the uncoupled coherent ferrons ($\omega_f$). The white horizontal lines indicate the resonant frequencies of the bulk acoustic phonons, $\omega_n$, with $n$=1,3,5 (if applicable) from bottom to top. (**b**, **d**) $E_3$-dependent frequency offset with respect to the uncoupled ferron, $\Delta\omega_f$, and $n$=1,3,5 mode phonon, $\Delta\omega_n$, where $E_3$ varies from -0.1 MV/cm to 0.1 MV/cm at 298 K.

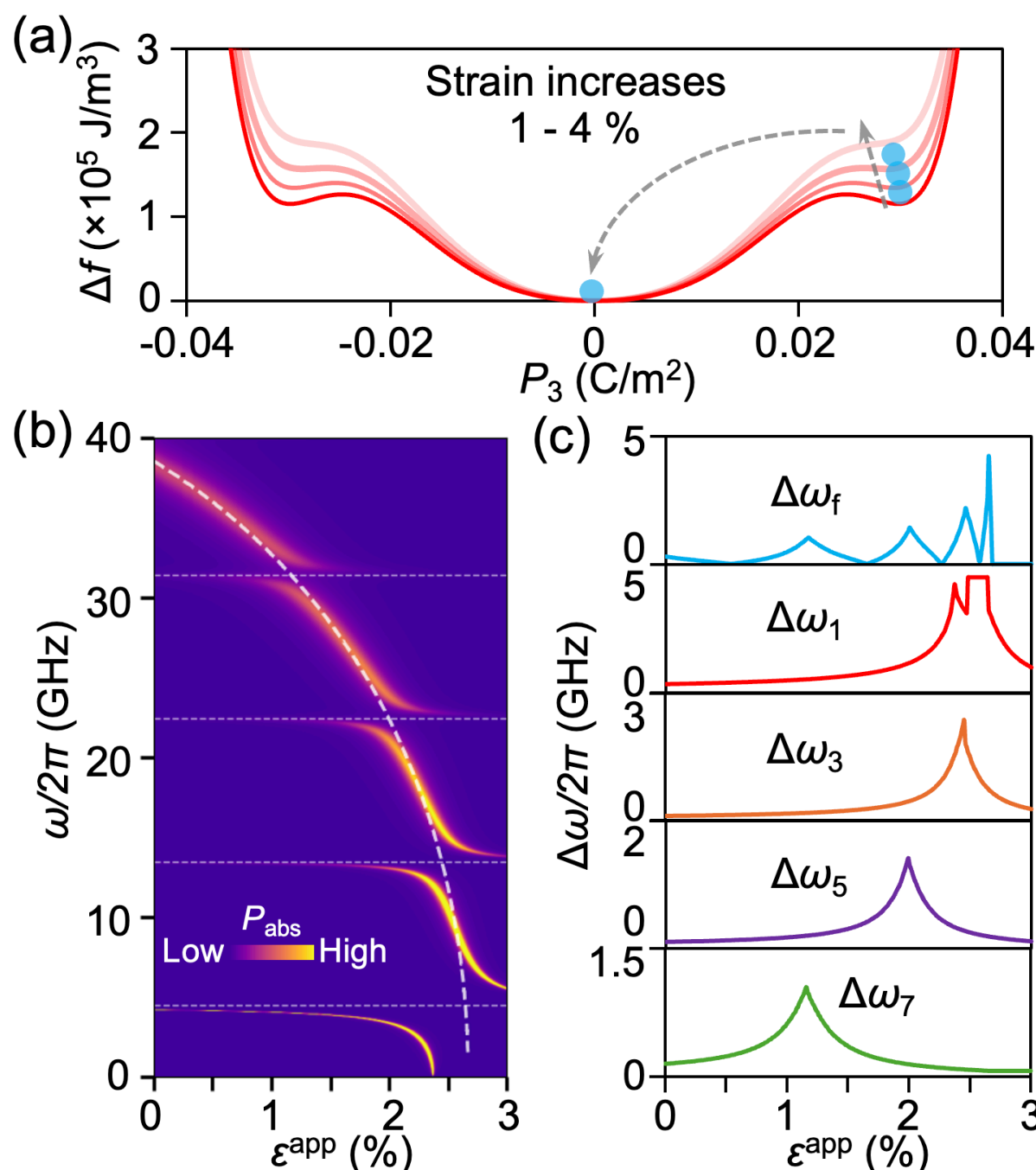

**Figure 4**. (**a**) Strain-dependent 1D free energy density $\Delta f$ in a uniaxially strained CIPS membrane at 305 K, where the applied strain $\varepsilon^{app}$=1%, 2%, 3%, and 4%. Blue dots indicate local energy minima. (**b**) Strain- and frequency-dependent power absorption spectrum of the driving microwave field, $P_{abs}$, in a 500-nm-thick CIPS membrane, where $\varepsilon^{app}$ varies from 0 to 3% at 305 K. The membrane would transform into a paraelectric phase when $\varepsilon^{app}$ exceeds 3.2%. The dashed curve and horizontal line indicate the $E_3$-dependent resonant frequencies of the uncoupled coherent ferrons, $\omega_f$, and $n$=1,3,5,7 mode bulk acoustic phonons, respectively. (**c**) Strain-dependent frequency offset with respect to the uncoupled ferron, $\Delta\omega_f$, and $n$=1,3,5,7 mode phonon, $\Delta\omega_n$.

## Supplemental Materials

**S1. The energy density function of $CuInP_2S_6$ and the refinement of relevant material parameters**

In the paraelectric phase, $CuInP_2S_6$ (CIPS) possesses a point group 2/*m* which is centrosymmetric [17]. Below the ferrielectric-to-paraelectric transition temperature (~312 K), the CIPS transforms into a monoclinic ferrielectric phase with a point group *m*. The ferrielectric CIPS possesses a robust net out-of-plane spontaneous polarization $P_3$ (along the $x_3$, or the *c* axis), which arises from the large upward displacement of $Cu^+$ ions (with respect to the midplane of the layer in the paraelectric 2/m phase) and much smaller downward shifting of the $In^{3+}$ ions [43]. By symmetry, the ferrielectric CIPS is allowed to develop a net in-plane spontaneous polarization $P_1$ along the $x_1$ (*a*) axis, but the magnitude of $P_1$ is likely negligible and was not detected experimentally [43]. The presence of predominant out-of-plane $P_3$ is supported by the first-principles density functional theory (DFT) calculations [41]. Therefore, we consider that the net spontaneous polarization in CIPS is effectively uniaxial and only has the out-of-plane $P_3$ component. Because such $P_3$ represents a collective polarization coordinate of the antiphase translation of $Cu^+$ and $In^{3+}$ ions along $x_3$ (i.e., the ferroelectric soft-mode illustrated in Fig. 1(a)), the resonant oscillation frequency of $P_3$ can be much lower than the frequency of the translational vibration of individual $Cu^+$ or $In^3$ ions (at around 1.5 THz in temporal frequency based on Raman spectroscopy [44]), especially when the coupling between these two polar sublattice is weak. In bulk CIPS crystal, dielectric spectroscopy [45] indicates that the resonant oscillation frequency of $P_3$, i.e., the resonant frequency of the ferroelectric soft mode (or fundamental-mode coherent ferron), reduces to the GHz level near the ferrielectric-to-paraelectric transition temperature. We now proceed to the derivation of the energy densities.

The Helmholtz free energy density of a single-domain ferroelectric material is given by $f^{\mathrm{E}}(T,P_i,E_i,\varepsilon_{ij})=g_0(T)+f^{\mathrm{Landau}}+f^{\mathrm{Elast}}+f^{\mathrm{Elec}}$ [46], where $T$, $P_i$, $E_i$, $\varepsilon_{ij}$ (strain) are the natural variables, $g_0(T)$ is the Gibbs free energy density of the centrosymmetric reference phase with zero spontaneous polarization, zero electric field, and zero stress.

For CIPS, $f^{\mathrm{Landau}}$ can be written as,

$$f^{\mathrm{Landau}} = \alpha_{33}(T)P_3^2 + \alpha_{3333}P_3^4 + \alpha_{333333}P_3^6 + \alpha_{33333333}P_3^8, \qquad (\mathrm{S1}-1)$$

where $\alpha_{11}$, $\alpha_{22}$, $\alpha_{33}$, $\alpha_{3333}$, $\alpha_{333333}$, and $\alpha_{33333333}$ are the Landau coefficients under stress-free condition. The elastic free energy density is given as $f^{\mathrm{Elast}} = \frac{1}{2}c_{ijkl}e_{kl}e_{ij}$, where the elastic strain $e_{ij} = \varepsilon_{ij} - \varepsilon_{ij}^0$, and the elastic stiffness tensor $c_{ijkl}$ has the symmetry of the 2/*m* phase for the CIPS. Accordingly, the $f^{\mathrm{Elast}}$ can be expanded as,

$$\begin{aligned} f^{\mathrm{Elast}} = {} & \frac{1}{2}c_{11}e_{11}^2 + \frac{1}{2}c_{22}e_{22}^2 + \frac{1}{2}c_{33}e_{33}^2 + c_{12}e_{11}e_{22} + c_{13}e_{11}e_{33} + c_{23}e_{22}e_{33} + 2c_{44}e_{23}^2 \\ & + 2c_{55}e_{13}^2 + 2c_{66}e_{12}^2 + 2c_{15}e_{11}e_{13} + 2c_{25}e_{22}e_{13} + 2c_{35}e_{33}e_{13} \\ & + 4c_{46}e_{23}e_{12}, \end{aligned} \qquad (\mathrm{S1}-2)$$

The eigenstrains $\varepsilon_{ij}^{0} = Q_{ijkl}P_kP_l$, where the electrostrictive coefficient tensor $Q_{ijkl}$ also has $2/m$ point group symmetry. Since the spontaneous polarization only has the $P_3$ component, $\varepsilon_{ij}^{0}$ can be expanded as,

$$\varepsilon_{11}^{0} = Q_{13}P_3^2,\ \ \varepsilon_{22}^{0} = Q_{23}P_3^2,\ \ \varepsilon_{33}^{0} = Q_{33}P_3^2,\ \ \varepsilon_{23}^{0} = 0,\ \ \varepsilon_{13}^{0} = Q_{53}P_3^2,\ \ \varepsilon_{12}^{0} = 0 \qquad (\mathrm{S1}-3)$$

At the equilibrium state, the total strain $\varepsilon_{ij}$ is determined by the mechanical boundary condition and the equilibrium polarization $P_i^{\mathrm{eq}}$. For a stress-free (unclamped) single-domain ferroelectric membrane, $\varepsilon_{ij}^{\mathrm{eq}} = \varepsilon_{ij}^{0}\left(P_i^{\mathrm{eq}}\right)$. At a nonequilibrium state, $f^{\mathrm{Elast}}$ is calculated under a fixed $\varepsilon_{ij} = \varepsilon_{ij}^{0}\left(P_i^{\mathrm{eq}}\right)$ and a time-varying $\varepsilon_{ij}^{0} = \varepsilon_{ij}^{0}(P_i)$.

The electrostatic energy density $f^{\mathrm{Elec}}$ is given as,

$$f^{\mathrm{Elec}}(\mathbf{P}, E_i) = -\frac{1}{2}D_iE_i = -\frac{1}{2}\left(\kappa_0\kappa_{\mathrm{b}}E_j + P_i\right)E_i = -\frac{1}{2}\kappa_0\kappa_{\mathrm{b}}E_jE_i - E_iP_i, \quad (\mathrm{S1}-4\mathrm{a})$$

$$E_i = E_i^{\mathrm{ext}} + E_i^{\mathrm{d}} = \left(E_i^{\mathrm{inc}} + E_i^{\mathrm{rad}}\right) + E_i^{\mathrm{d}}, \qquad (\mathrm{S1}-4\mathrm{b})$$

where $\kappa_0$ is the vacuum permittivity and $\kappa_{\mathrm{b}}$ is the background permittivity. $E_i^{\mathrm{inc}}$ is the electric field component of the incident microwave field and $E_i^{\mathrm{d}}$ is the depolarization field. By assuming a complete screening of the polarization charges at the top and bottom surfaces of a CIPS membrane by mobile charges, $E_i^{\mathrm{d}}$ at $P$=$P_3^{\mathrm{eq}}$ is zero. This assumption is reasonable because an out-of-plane (along $x_3$) spontaneous polarization has recently been experimentally observed in thin CIPS membrane (down to ~8 nm) [47].

We further note that the ferroelectric soft mode in CIPS can be approximately considered as a transverse optical (TO) phonon, represented by $\Delta P_3(k^{(1)},\omega)$ or $\Delta P_3(k^{(2)},\omega)$, where $k^{(1)}$ and $k^{(2)}$ are the wavenumbers along the $x_1$ and $x_2$ axis, respectively, and the 'Δ' quantifies the change with respect to $P_3^{\mathrm{eq}}$. It is relatively straightforward to prove that these TO phonons would not cause significant out-of-plane dynamical depolarization field, i.e., $\Delta E_3^{\mathrm{d}}(t)\approx 0$. Furthermore, the out-of-plane radiation field component $E_3^{\mathrm{rad}}$ is also zero because the TO phonons, $\Delta P_3(k^{(1)},\omega)$ or $\Delta P_3(k^{(2)},\omega)$, do not generate a radiative electric field along the $x_3$ axis. The in-plane components $E_1^{\mathrm{rad}}$ and $E_2^{\mathrm{rad}}$ should be nonzero and affect the attenuation of $\Delta P_3(k^{(1)},\omega)$ or $\Delta P_3(k^{(2)},\omega)$, but such an effect is not related to the scope of this work.

The gradient energy density is calculated as $f^{\mathrm{Grad}} = \frac{1}{2}G_{ijkl}\frac{\partial P_i}{\partial x_j}\frac{\partial P_k}{\partial x_l}$, where the gradient energy coefficient tensor $G_{ijkl}$ likewise has the symmetry of $2/m$ phase. $f^{\mathrm{Grad}}$ can be expanded into,

$$f^{\mathrm{Grad}} = \frac{1}{2}G_{3131}\left(\frac{\partial P_3}{\partial x_1}\right)^2 + \frac{1}{2}G_{3232}\left(\frac{\partial P_3}{\partial x_2}\right)^2 + \frac{1}{2}G_{3333}\left(\frac{\partial P_3}{\partial x_3}\right)^2 + G_{3133}\frac{\partial P_3}{\partial x_1}\frac{\partial P_3}{\partial x_3}. \quad (\mathrm{S1}-5)$$

These gradient energy coefficients are estimated to be on the order of $10^{-10}$ J m$^3$/C$^2$ [17]. If further omitting the possible spatial variation of $P_3$ along $x_1$ and $x_2$, i.e. $\partial P_3/\partial x_1$=$\partial P_3/\partial x_2$=0, one has $f^{\mathrm{Grad}} \approx \frac{1}{2} G_0 \left(\frac{\partial P_3}{\partial x_3}\right)^2$, where $G_0 \equiv G_{3333}$. A large gradient energy coefficient $G_0$ will therefore surpresss the formation of nonuniform polarization wave along the $x_3$ axis.

Next, we discuss the refinement of the material parameters of CIPS used in the analytical calculation and dynamical phase-field simulations. Table S1 lists all the parameters we used and their sources. Below, we discuss the determination of $c_{33}$ and $Q_{33}$ based on the experimentally measured temperature-dependent ultrasonic velocity [45].

It is worth remarking that the elastic stiffness tensor components $c_{ijkl}$ in Eq. (S1-2) are determined under zero spontaneous polarization, which is a typical treatment in Landau-Ginzburg-Devonshire (LGD) theory [48,49], and hence have the symmetry of the parent paraelectric phase ($2/m$). However, the experimentally measured sound velocities in the ferrielectric phase are determined by the effective (constant electric-field) elastic stiffness tensor $c_{ijkl}^{\mathrm{E}}$, which is further related to the electrostrictive tensor $Q_{ijkl}$ and the piezoelectric coupling, as will be elaborated below.

Let us consider the ultrasonic measurement reported in [45], where a low-frequency acoustic wave (10 MHz, much lower than the GHz resonant oscillation frequency of the spontaneous polarization $P_n$) is injected into the CIPS crystal via a juxtaposed piezoelectric transducer. Under such a MHz acoustic wave, $P_n$ has sufficient time to evolve to its new thermodynamic equilibrium before the strain changes. We further assume a spatially uniform ground-state $P_n$ in the CIPS crystal and the polarization wave excited by the 10 MHz acoustic wave has a long wavelength. Under this condition, one can write down the equations of state,

$$\sigma_{ij} = \left(\frac{\partial f^{\mathrm{E}}}{\partial \varepsilon_{ij}}\right)_{T,P_i,E_i}, \qquad c_{ijkl}^{\mathrm{E}} = \frac{\partial \sigma_{ij}}{\partial \varepsilon_{kl}} = \left(\frac{\partial^2 f^{\mathrm{E}}}{\partial \varepsilon_{kl} \partial \varepsilon_{ij}}\right)_{T,P_i,E_i}. \qquad (\mathrm{S1-6})$$

Given that $\sigma_{ij} = c_{ijkl}(\varepsilon_{kl} - \varepsilon_{kl}^0) = c_{ijkl}\varepsilon_{kl} - c_{ijkl}Q_{klmn}P_m P_n$, and considering the symmetry of the electrostrictive tensor (i.e., $Q_{klmn} = Q_{klnm}$), $c_{ijkl}^{\mathrm{E}}$ can be expanded into,

$$c_{ijkl}^{\mathrm{E}} = c_{ijkl} - 2c_{ijkl}Q_{klmn}P_m \frac{\partial P_n}{\partial \varepsilon_{kl}}, \qquad (\mathrm{S1-7})$$

where $d_{nkl} = \partial P_n / \partial \varepsilon_{kl}$ is the electromechanical coupling tensor. $d_{nkl}$ and the resulting $c_{ijkl}^{\mathrm{E}}$ can be further derived by the minimization of $f^{\mathrm{E}}$ with respect to the polarization, i.e., $\partial f^{\mathrm{E}} / \partial P_i$=0. By further differentiating the term $(\partial f^{\mathrm{E}} / \partial P_i)$ with respect to the total strain $\varepsilon_{kl}$, applying the chain rule, and noting that both $P_i$ and $\varepsilon_{kl}$ are state variables, one can write,

$$\frac{\partial}{\partial \varepsilon_{kl}}\left(\frac{\partial f^{\mathrm{E}}}{\partial P_i}\right) = \frac{\partial^2 f^{\mathrm{E}}}{\partial P_i \partial P_n}\frac{\partial P_n}{\partial \varepsilon_{kl}} + \frac{\partial^2 f^{\mathrm{E}}}{\partial P_i \partial \varepsilon_{kl}} = K_{in}\frac{\partial P_n}{\partial \varepsilon_{kl}} + L_{ikl} = 0, \qquad (\mathrm{S1-8})$$

where $K_{in} = \partial^2 f^{\mathrm{E}} / (\partial P_i \partial P_n)$, and $L_{ikl} = \partial^2 f^{\mathrm{E}} / (\partial P_i \partial \varepsilon_{kl})$. Because the coupling between the polarization and strain is described by the elastic energy density $f^{\mathrm{Elast}}$, $L_{ikl}$ can be further written as $L_{ikl} = \partial^2 f^{\mathrm{E}} / (\partial P_i \partial \varepsilon_{kl}) = \partial^2 f^{\mathrm{Elast}} / (\partial P_i \partial \varepsilon_{kl}) = -2c_{klrs}Q_{rsiq}P_q$.

Combining Eq.(S1-7,8) allows for deriving $c_{ijkl}^{\mathrm{E}}$ related to low-frequency ultrasonic measurement,

$$c_{ijkl}^{\mathrm{E}} = c_{ijkl} - 4c_{ijkl}Q_{klmn}P_m K_{in}^{-1} c_{klrs}Q_{rsiq}P_q, \qquad (\mathrm{S1}-9)$$

The ultrasonic measurement in [45] involves the propagation of longitudinal acoustic (LA) phonons along $x_3$, which is related to the longitudinal speed of sound $v = \sqrt{c_{33}^{\mathrm{E}}/\rho}$, where $\rho$ is the mass density. Based on Eq. (S1-9), dropping smaller terms, $c_{33}^{\mathrm{E}}$ can be expressed as,

$$c_{33}^{\mathrm{E}} \approx c_{33} - \frac{4{P_3^{\mathrm{eq}}}^2 c_{33}^2 Q_{33}^2}{K_{33}}. \qquad (\mathrm{S1}-10)$$

We note that $P_3^{\mathrm{eq}}$ is a function of temperature $T$ (see Fig. 1(d)), which can be obtained by the minimization of $f^{\mathrm{E}}$. $K_{33}$, which describes the curvature of the free energy landscape at $P_3$=$P_3^{\mathrm{eq}}$, is a function of both $c_{33}$ and $Q_{33}$. As shown in Fig. 1(d), $K_{33}$ is always positive. We can therefore calculate the $v(T)$ of the low-frequency acoustic wave based on Eq. (S1-10). By comparing the analytically calculated and experimentally measured $v(T)$, the values of $c_{33}$ and $Q_{33}$ are refined. The refined values are listed in Table S1.

Figure S1 shows the analytically calculated and experimentally measured $v(T)$, which agrees well with each other in the ferrielectric phase. Specifically, when approaching the transition temperature (~312 K) from the ferrielectric phase, the significantly reduced local curvature leads to a substantially reduced $c_{33}^{\mathrm{E}}$ and hence the sound velocity. At near and above the ferrielectric-to-paraelectric transition temperature (~312 K), however, the present Landau theory-based approach fails to describe the $v(T)$. This discrepancy may result from the spatially inhomogeneous ground-state polarization configuration that is believed to emerge near the ferrielectric-to-paraelectric phase transition temperature in bulk CIPS crystal (see discussion in [50]), yet the Landau theory typically assumes a homogeneous ground-state polarization configuration.

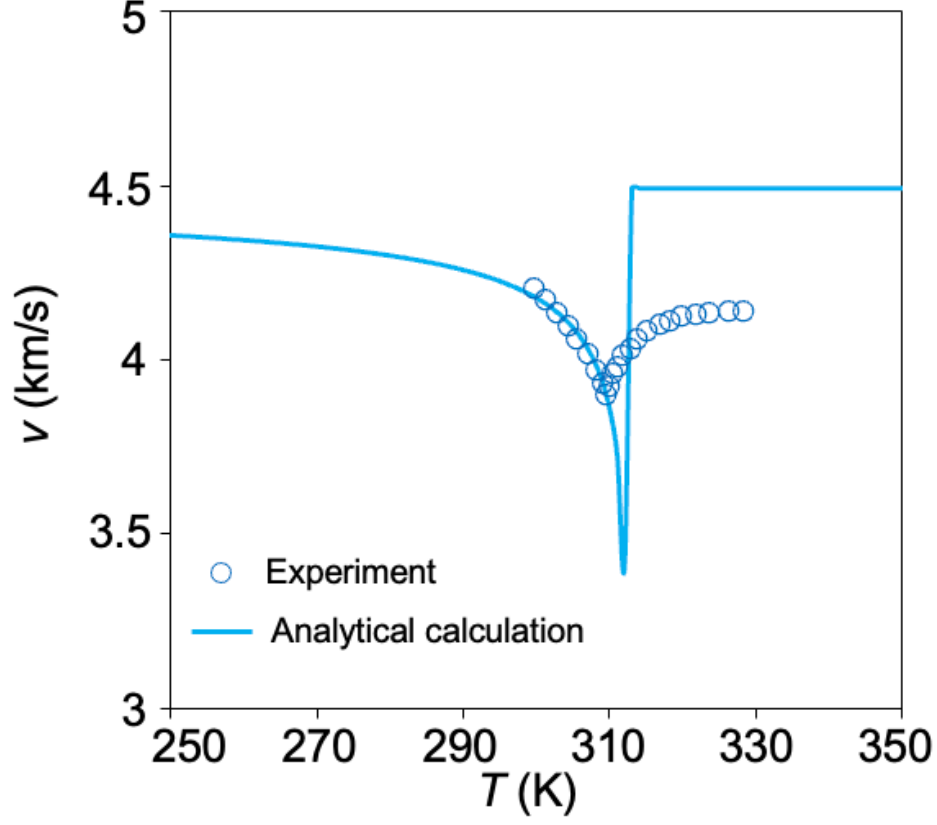

**Figure S1.** Temperature-dependent longitudinal sound velocity in bulk CIPS bulk crystals, corresponding to the propagation of 10 MHz LA phonons along the $x_3$ axis. The analytical calculation is performed by using $v = \sqrt{c_{33}^{\mathrm{E}}/\rho}$ (the expression of $c_{33}^{\mathrm{E}}$ is in Eq. (S1-10)) to fit the experimental data below the ferrielectric-to-paraelectric transition temperature (~312 K).

**Table S1.** List of the coefficients of the $CuInP_2S_6$

| | Coefficients | Value |
|---|---|---|
| Landau coefficients | $\alpha_{ij}$ (N m$^2$ C$^{-2}$) | $\alpha_{33}(T)$=8.20335×10$^6$($T$-283.17)[(a)]. |
| | $\alpha_{ijkl}$ (N m$^6$ C$^{-4}$) | $\alpha_{3333}$=7.87×10$^{11}$ [51]. |
| | $\alpha_{ijklmn}$ (N m$^{10}$ C$^{-6}$) | $\alpha_{333333}$=-1.796×10$^{15}$ [51]. |
| | $\alpha_{ijklmnpq}$ (N m$^{14}$ C$^{-8}$) | $\alpha_{33333333}$=9.53975×10$^{17}$ [51]. |
| Elastic stiffness tensor | $\begin{pmatrix} c_{11} & c_{12} & c_{13} & 0 & c_{15} & 0 \\ c_{12} & c_{22} & c_{23} & 0 & c_{25} & 0 \\ c_{13} & c_{23} & c_{33} & 0 & c_{35} & 0 \\ 0 & 0 & 0 & c_{44} & 0 & c_{46} \\ c_{15} & c_{25} & c_{35} & 0 & c_{55} & 0 \\ 0 & 0 & 0 & c_{46} & 0 & c_{66} \end{pmatrix}$ (GPa) | $\begin{pmatrix} 99.86 & 29.01 & -0.86 & 0 & 0.17 & 0 \\ 29.01 & 101.7 & -1.93 & 0 & 0.94 & 0 \\ -0.86 & -1.93 & 69.146 & 0 & -0.62 & 0 \\ 0 & 0 & 0 & 6.99 & 0 & 0.69 \\ 0.17 & 0.94 & -0.62 & 0 & 6.71 & 0 \\ 0 & 0 & 0 & 0.69 & 0 & 37.56 \end{pmatrix}$[(b)] |
| Electrostrictive tensor | $\begin{pmatrix} Q_{11} & Q_{12} & Q_{13} & 0 & Q_{15} & 0 \\ Q_{21} & Q_{22} & Q_{23} & 0 & Q_{25} & 0 \\ Q_{31} & Q_{32} & Q_{33} & 0 & Q_{35} & 0 \\ 0 & 0 & 0 & Q_{44} & 0 & Q_{46} \\ Q_{51} & Q_{52} & Q_{53} & 0 & Q_{55} & 0 \\ 0 & 0 & 0 & Q_{64} & 0 & Q_{66} \end{pmatrix}$ (m$^4$ C$^{-2}$) | $\begin{pmatrix} 0 & 0 & 1.70136 - 0.00363T & 0 & 0.1 & 0 \\ 0 & 0 & 1.13424 - 0.00242T & 0 & 0.1 & 0 \\ 0 & 0 & -1.7858 & 0 & 0.1 & 0 \\ 0 & 0 & 0 & 0.1 & 0 & 0 \\ 0 & 0 & 0.1 & 0 & 0.1 & 0 \\ 0 & 0 & 0 & 0.1 & 0 & 0 \end{pmatrix}$[(c)] |
| Effective mass coefficient | $\mu$ (J m s$^2$ C$^{-2}$) | 8×10$^{-14}$ [17] |
| Polarization damping | $\gamma$ (Ω·m) | 10$^{-3}$ [17] |
| Background permittivity | $\kappa_b$ | 9 [17] |
| Mass density | $\rho$ (kg/m$^3$) | 3427 [17] |
| Elastic damping | $\beta$ (s) | 9.19×10$^{-14}$ [(d)] |

(a) The expression of $\alpha_{33}$ was adjusted from that provided in [51] to ensure the simulated ferrielectric-to-paraelectric phase transition temperature is consistent with the experimentally measured value of 312 K [43]. The temperature $T$ is in K.

(b) The values of $c_{11}$, $c_{22}$, $c_{44}$, $c_{55}$, and $c_{66}$ are available in [52]. Based on these values, the values of $c_{12}$, $c_{13}$, and $c_{23}$ are further evaluated based on the Poisson's ratio of CIPS, given in [53]. The determination of $c_{33}$ is discussed in the text above. The remaining components ($c_{15}$, $c_{25}$, $c_{35}$, and $c_{46}$) are taken from [47].

(c) The determination of $Q_{33}$ is discussed in the text above. The remaining components of the electrostrictive tensor are available in [47].

(d) The determination of the elastic damping coefficient is discussed in Sec. S4.

## S2. Derivation of dynamic dielectric and electromechanical susceptibilities

We consider the CIPS membrane structure to be a 1D system where the physical quantities are uniform in the $x_1$-$x_2$ plain and only vary along the $x_3$ direction from $x_3$=0 to $x_3$=$d$. CIPS has a spontaneous polarization along the $x_3$ axis at thermodynamic equilibrium, i.e. $\mathbf{P}^{\mathrm{eq}} = \left(0,0,P_3^{\mathrm{eq}}\right)$, and the spontaneous polarization value $P_3^{\mathrm{eq}}$ can be estimated by minimizing the Helmholtz free energy density $f^{\mathrm{E}}$ with respect to $\mathbf{P}$. An incident c.w. microwave electric field $\mathbf{E}^{\mathrm{inc}}(\omega) = \left(0,0,E_3^0 e^{-\mathrm{i}\omega t}\right)$ is applied to the system. As discussed in the main text (Fig. 1c), we consider a relatively large gradient coefficient $G_0$ to suppresses the $k$≠0 mode ferron, such that the strong coupling can primarily occur between the fundamental-mode ($k$=0) ferron and bulk cavity acoustic phonons. Considering only the harmonic polarization oscillation, the lattice polarization in CIPS can be written as $\mathbf{P}(x_3,\omega) = \mathbf{P}^{\mathrm{eq}} + \Delta\mathbf{P}(\omega) = \left(\Delta P_1^0 e^{-\mathrm{i}\omega t}, \Delta P_2^0 e^{-\mathrm{i}\omega t}, P_3^{\mathrm{eq}} + \Delta P_3^0 e^{-\mathrm{i}\omega t}\right)$, where the complex amplitude can be written as $\Delta P_i^0 = \left|\Delta P_i^0\right| e^{\mathrm{i}\theta_i^{\mathrm{P}}}$, $\left|\Delta P_i^0\right|$ denotes the oscillation amplitude of the polarization oscillation, and $\theta_i^{\mathrm{P}}$ denotes the phase angle relative to $E_3^{\mathrm{inc}}$. Likewise, we write the mechanical displacement $u_i(x_3,\omega) = u_i^{\mathrm{eq}} + \Delta u_i^0(x_3) e^{-\mathrm{i}\omega t}$ , strain $\varepsilon_{ij}(x_3,\omega) = \varepsilon_{ij}^{\mathrm{eq}} + \Delta\varepsilon_{ij}^0(x_3) e^{-\mathrm{i}\omega t}$, where $\Delta u_i^0(x_3)$ and $\Delta\varepsilon_{ij}^0(x_3)$ are both complex values with a phase angle relative to $E_3^{\mathrm{inc}}$. The complex-valued dielectric susceptibility tensor is defined by, $\chi_{ij}(\omega) \approx \frac{\Delta P_i}{\kappa_0 E_j^{\mathrm{inc}}} = \frac{\Delta P_i^0}{\kappa_0 E_j^0}$. $\chi_{ij}(\omega)$ can be derived by linearizing the coupled equations of motion for polarization and mechanical displacement, given by [32,59,60],

$$\mu \frac{\partial^2 P_i}{\partial t^2} + \gamma \frac{\partial P_i}{\partial t} = -\frac{\delta F}{\delta P_i}, \qquad \text{(S2 − 1)}$$

$$\rho \frac{\partial^2 u_i}{\partial t^2} = \nabla \cdot \left(\sigma_{ij} + \beta \frac{\partial \sigma_{ij}}{\partial t}\right). \qquad \text{(S2 − 2)}$$

Here, $\mu$ is the effective mass coefficient related to the ionic mass and Born effective charge of an unit cell, $\gamma$ and $\beta$ are the phenomenological damping coefficients. Stress is calculated via $\sigma_{ij} = c_{ijkl}\left(\varepsilon_{kl} - Q_{ijkl} P_k P_l\right)$ , where the total strain tensor, $\varepsilon_{kl} = \frac{1}{2}\left(\frac{\partial u_k}{\partial x_l} + \frac{\partial u_l}{\partial x_k}\right)$ describing the deformation with respect to the parent paraelectric phase. $F = \int\left(f^{\mathrm{E}} + f^{\mathrm{Grad}}\right) dV$ is the total free energy of the ferroelectric. The expressions of $f^{\mathrm{E}}$ and $f^{\mathrm{Grad}}$ as well as the relevant material parameters are provided in Sec. S1.

Substituting the expressions of $\mathbf{P}(x_3,\omega)$ and $\varepsilon_{ij}(x_3,\omega)$ into Eq. (S2-1), neglecting the higher-order terms, and assuming the amplitude of polarization oscillation is spatially uniform along $x_3$ (which is the case for $k$=0 mode coherent ferron), Eq. (S2-1) can be rewritten as,

$$\begin{pmatrix} A_{11} & 0 & 0 \\ 0 & A_{22} & 0 \\ 0 & 0 & A_{33} \end{pmatrix} \begin{pmatrix} \Delta P_1^0 \\ \Delta P_2^0 \\ \Delta P_3^0 \end{pmatrix} e^{-\mathrm{i}\omega t} + \begin{pmatrix} L_{133} & 0 & L_{113} \\ 0 & L_{223} & 0 \\ L_{333} & 0 & L_{313} \end{pmatrix} \begin{pmatrix} \Delta\varepsilon_{33}^0 \\ \Delta\varepsilon_{23}^0 \\ \Delta\varepsilon_{13}^0 \end{pmatrix} e^{-\mathrm{i}\omega t} = \begin{pmatrix} 0 \\ 0 \\ E_3^0 \end{pmatrix} e^{-\mathrm{i}\omega t}, \qquad \text{(S2 − 3)}$$

where $A_{ii} = \mu\left(\omega_i^{\mathrm{f}^2} - \omega^2\right) - \mathbf{i}\gamma\omega$. $\omega_i^{\mathrm{f}}$ denotes the resonant frequency of $P_i$, with $\omega_i^{\mathrm{f}}=\sqrt{K_{ii}/\mu}$. The definitions of $K_{ii}$ and $L_{ijk}$ are provided in Sect. S1, below Eq. (S1-8).

Next, we take a spatial average for the terms on both sides of Eq. (S2-3) along the thickness direction, i.e. $\langle\Theta\rangle \equiv \frac{1}{d}\int_0^d \left(\Theta(x_3)\right) dx_3$. Since $\Delta P_i^0$ and $E_3^0$ are both spatially uniform, i.e. $\langle\Delta P_i^0\rangle = \Delta P_i^0$, $\langle E_3^0\rangle = E_3^0$, Eq. (S2-3) can be rewritten as,

$$\begin{pmatrix} A_{11} & 0 & 0 \\ 0 & A_{22} & 0 \\ 0 & 0 & A_{33} \end{pmatrix}\begin{pmatrix} \Delta P_1^0 \\ \Delta P_2^0 \\ \Delta P_3^0 \end{pmatrix} + \begin{pmatrix} L_{133} & 0 & L_{113} \\ 0 & L_{223} & 0 \\ L_{333} & 0 & L_{313} \end{pmatrix}\begin{pmatrix} \langle\Delta\varepsilon_{33}^0\rangle \\ \langle\Delta\varepsilon_{23}^0\rangle \\ \langle\Delta\varepsilon_{13}^0\rangle \end{pmatrix} = \begin{pmatrix} 0 \\ 0 \\ E_3^0 \end{pmatrix}, \quad \text{(S2-4)}$$

As shown in Eq. (S2-4), $\Delta P_2$ cannot be excited, with $\Delta P_2$=0. $\Delta P_3$ can be excited directly by $E_3^{\mathrm{inc}}$ via $\chi_{33}$. $\langle\Delta\varepsilon_{33}\rangle$ and $\langle\Delta\varepsilon_{13}\rangle$ can be excited by $E_3^{\mathrm{inc}}$ via $L_{333}$ and $L_{313}$, respectively, although the magnitude of $\langle\Delta\varepsilon_{13}\rangle$ will be smaller because $|L_{313}|<|L_{333}|$. Once $\langle\Delta\varepsilon_{33}\rangle$ and $\langle\Delta\varepsilon_{13}\rangle$ are excited, they can further excite $\Delta P_1$ via $L_{133}$ and $L_{113}$, respectively. However, $\Delta P_1$ excited through such secondary effect would be negligibly small compared to $\Delta P_3$, because $L_{133}$ and $L_{113}$ are significantly smaller than $L_{333}$ and $L_{313}$. For example, at 298 K, we obtain $P_3^{\mathrm{eq}}$=0.0305 C/m$^2$ by minimizing $f^{\mathrm{E}}$, yielding $L_{133}$=-199.0 MV/m, $L_{113}$=-85.0 MV/m, $L_{333}$=3418.3 MV/m, $L_{313}$=-1069.4 MV/m. Therefore, we assume $\Delta P_1 = 0$. Taken together, Eq. (S2-4) reduces to,

$$A_{33}\Delta P_3^0 + L_{333}\langle\Delta\varepsilon_{33}^0\rangle + L_{313}\langle\Delta\varepsilon_{13}^0\rangle = E_3^0, \quad \text{(S2-5)}$$

where $A_{33} = \mu\left(\omega_{\mathrm{f}}^2 - \omega^2\right) - \mathbf{i}\gamma\omega$, $\omega_{\mathbf{f}} \equiv \omega_3^{\mathrm{f}}$. For linear excitation, the relationship between the dynamic strain $\langle\Delta\varepsilon_{i3}\rangle$=$\langle\Delta\varepsilon_{i3}^0\rangle e^{-\mathbf{i}\omega t}$($i$=1,3) and $\Delta P_3$=$\Delta P_3^0 e^{-\mathbf{i}\omega t}$, should also be linear, given by

$$\langle\Delta\varepsilon_{33}^0\rangle(\omega) = \Omega_{333}(\omega)\Delta P_3^0, \quad \text{(S2-6a)}$$

$$\langle\Delta\varepsilon_{13}^0\rangle(\omega) = \Omega_{313}(\omega)\Delta P_3^0, \quad \text{(S2-6b)}$$

where $\Omega_{313}(\omega)$ and $\Omega_{333}(\omega)$ are frequency-dependent electromechanical coupling coefficients. Substituting Eqs. (S2-6a,b) into Eq. (S2-5), $\chi_{33}(\omega)$ can be written as,

$$\chi_{33}(\omega) = \frac{1}{\kappa_0}\frac{\Delta P_3^0}{E_3^0} = \frac{1}{\kappa_0}\frac{1}{\mu(\omega_{\mathrm{f}}^2 - \omega^2) - \mathbf{i}\gamma\omega + L_{333}\Omega_{313} + L_{313}\Omega_{333}}, \quad \text{(S2-7)}$$

Given that $|L_{313}| \ll |L_{333}|$, Eq. (S2-7) reduces to,

$$\chi_{33}(\omega) \approx \frac{1}{\kappa_0}\frac{\Delta P_3^0}{E_3^0} = \frac{1}{\kappa_0}\frac{1}{\mu(\omega_{\mathrm{f}}^2 - \omega^2) - \mathbf{i}\gamma\omega + L_{333}\Omega_{333}(\omega)}, \quad \text{(S2-8)}$$

which is Eq. (1) in the main text.

Next, we derive the expression of $\Omega_{313}(\omega)$ and $\Omega_{333}(\omega)$ by linearizing the elastodynamic equation [Eq. (S2-2)], as discussed below.

By substituting $u_i(x_3,t) = u_i^{\mathrm{eq}} + \Delta u_i(x_3,t)$ into Eq. (S2-2), and omitting the coupling between $\Delta u_1$ and $\Delta u_3$ via $c_{35}$ (which is one-to-two order of magnitude smaller than the diagonal components of elastic stiffness tensor $c_{33}$ and $c_{55}$, see Table S1), Eq. (S2-2) can be expanded into,

$$\rho\frac{\partial^2\Delta u_1(x_3,t)}{\partial t^2} - c_{55}\frac{\partial}{\partial x_3^2}\left(1+\beta\frac{\partial}{\partial t}\right)\Delta u_1(x_3,t) = \frac{L_{313}}{2}\frac{\partial}{\partial x_3}\left(1+\beta\frac{\partial}{\partial t}\right)\Delta P_3(x_3,t), \quad \text{(S2 − 9a)}$$

$$\rho\frac{\partial^2\Delta u_2(x_3,t)}{\partial t^2} - c_{44}\frac{\partial}{\partial x_3^2}\left(1+\beta\frac{\partial}{\partial t}\right)\Delta u_2(x_3,t) = 0, \quad \text{(S2 − 9b)}$$

$$\rho\frac{\partial^2\Delta u_3(x_3,t)}{\partial t^2} - c_{33}\frac{\partial}{\partial x_3^2}\left(1+\beta\frac{\partial}{\partial t}\right)\Delta u_3(x_3,t) = L_{333}\frac{\partial}{\partial x_3}\left(1+\beta\frac{\partial}{\partial t}\right)\Delta P_3(x_3,t), \quad \text{(S2 − 9c)}$$

Equations (S2-8) indicate that the $\Delta P_3(x_3,t)$ is coupled to $\Delta u_1(x_3,t)$ and $\Delta u_3(x_3,t)$ via $L_{313}$ and $L_{333}$, respectively. These couplings will lead to an effective elastic stiffness coefficient ($c_{33}^{\mathrm{eff}}$ or $c_{55}^{\mathrm{eff}}$) that depends on both the frequency and wavenumber of the acoustic phonons. This study is focused on the fundamental-mode coherent ferron, in which case $\frac{\partial \Delta P_3(t)}{\partial x_3}$=0. As a result, Eqs. (S2-9a,b,c) reduce to,

$$\rho\frac{\partial^2\Delta u_1(x_3,t)}{\partial t^2} - c_{55}\frac{\partial}{\partial x_3^2}\left(1+\beta\frac{\partial}{\partial t}\right)\Delta u_1(x_3,t) = 0, \quad \text{(S2 − 10a)}$$

$$\rho\frac{\partial^2\Delta u_2(x_3,t)}{\partial t^2} - c_{44}\frac{\partial}{\partial x_3^2}\left(1+\beta\frac{\partial}{\partial t}\right)\Delta u_2(x_3,t) = 0, \quad \text{(S2 − 10b)}$$

$$\rho\frac{\partial^2\Delta u_3(x_3,t)}{\partial t^2} - c_{33}\frac{\partial}{\partial x_3^2}\left(1+\beta\frac{\partial}{\partial t}\right)\Delta u_3(x_3,t) = 0, \quad \text{(S2 − 10c)}$$

Equations (S2-9) indicate that the sound velocities along the $x_3$ axis are determined by the bare elastic constants, specifically, $v = \sqrt{c_{33}/\rho}$ for the propagation of LA phonons, and $v = \sqrt{c_{55}/\rho}$ or $v = \sqrt{c_{44}/\rho}$ for the propagation of TA phonons. For cavity bulk acoustic wave (BAW) phonons, solutions to Eqs. (S2-9) should take the form,

$$\Delta u_i(x_3,t) = \Delta u_i^{+} e^{\mathbf{i}(k^{(i)}x_3-\omega t)} + \Delta u_i^{-} e^{-\mathbf{i}(k^{(i)}x_3+\omega t)}, \quad \text{(S2 − 11)}$$

where $\Delta u_i^{\pm}$ ($i$=1,2,3) are the amplitudes of the forward-propagating (along $+x_3$) and the backward-propagating (along $-x_3$) acoustic wave in the CIPS membrane. $k^{(i)}$ ($i$=1,2,3) are the wavenumbers of these acoustic waves, including $k^{(1)}$ and $k^{(2)}$ for TA phonons $\Delta u_1(x_3,t)$ and $\Delta u_2(x_3,t)$, respectively, as well as $k^{(3)}$ for LA phonons $\Delta u_3(x_3,t)$. The expressions of $k^{(i)}$ as a function of $\omega$ (the dispersion relation) can be determined by linearizing Eqs. (S2-9a,b,c). The expressions of $\Delta u_i^{\pm}$ can be obtained from the traction-free boundary condition at the top and bottom surfaces, $\Delta\sigma_{i3}(x_3 = 0,t) = \Delta\sigma_{i3}(x_3 = d,t) = 0$, because $\Delta\sigma_{i3}(x_3,t)$ is connected to both $\Delta u_i(x_3,t)$ (the spatial gradient of $\Delta u_i(x_3,t)$ corresponds to the dynamical total strain $\Delta\varepsilon_{i3}$) and $\Delta P_i(x_3,t)$ (which influences the dynamical eigenstrain $\Delta\varepsilon_{i3}^{0}$).

A knowledge of $\Delta u_i(x_3,t)$ allows for deriving the expressions of $\langle\Delta\varepsilon_{i3}\rangle$ as a function of $\Delta P_3$, which in turn yield the expressions of $\Omega_{333}$ and $\Omega_{313}$ (see Eqs. (2-6a,b)), given by,

$$\Omega_{313}(\omega) = -\frac{L_{313}}{d\omega\sqrt{c_{55}\rho}(-2\mathbf{i}+\beta\omega)}\tanh\left(\frac{d\omega\sqrt{\rho}(-2\mathbf{i}+\beta\omega)}{4\sqrt{c_{55}}}\right), \qquad (\mathrm{S2-12a})$$

$$\Omega_{333}(\omega) = -\frac{4L_{333}}{d\omega\sqrt{c_{33}\rho}(-2\mathbf{i}+\beta\omega)}\tanh\left(\frac{d\omega\sqrt{\rho}(-2\mathbf{i}+\beta\omega)}{4\sqrt{c_{33}}}\right), \qquad (\mathrm{S2-12b})$$

## S3. Frequency-dependent elastic stiffness coefficient

The dynamic coupling between $k \neq 0$ mode coherent ferrons, represented by the spatially nonuniform polarization oscillation $\Delta P_3(x_3, t)$, and the $k \neq 0$ mode acoustic phonons, represented by the oscillating mechanical displacement $\Delta u_3(x_3, t)$ will lead to a frequency-dependent effective elastic stiffness $c_{33}^{\text{eff}}(k, \omega)$, as indicated by Eq. (S2-9c).

We further consider the nonuniform $\Delta P_3(x_3, t)$ linearly excited by $\Delta u_3(x_3, t)$ via dynamical piezoelectric coupling (related to $\Omega_{333}(\omega)$ discussed in S2), hence $\Delta P_3(x_3, t)$ and $\Delta u_3(x_3, t)$ have the same frequency $\omega$ and wavenumber $k$, i.e. $\Delta P_3(x_3, t) = \Delta P_3^0 e^{\mathbf{i}(kx_3 - \omega t)}$, $\Delta u_3(x_3, t) = \Delta u_3^0 e^{\mathbf{i}(kx_3 - \omega t)}$, $k$=Re($k$)+ $\mathbf{i}$Im($k$) where Re($k$) and Im($k$) refer to the real and imaginary part of the wavenumber k, respectively. Substituting these into Eq.(S2-9c), the latter can be rewritten into,

$$\rho\omega^2 = k^2 c_{33}^{\text{tot}} = k^2(1 - \mathbf{i}\beta\omega)c_{33}^{\text{eff}}, \quad \text{(S3 - 1)}$$

where $c_{33}^{\text{tot}}$=$(1 - \mathbf{i}\beta\omega)c_{33}^{\text{eff}}$, and,

$$c_{33}^{\text{eff}} = c_{33} - \frac{\mathbf{i}L_{333}}{k}\frac{\Delta P_3^0}{\Delta u_3^0}, \quad \text{(S3 - 2)}$$

The ratio, $\Delta P_3^0/\Delta u_3^0$, can be obtained by substituting the expressions of $\Delta P_3(x_3, t)$ and $\Delta u_3(x_3, t)$ mentioned above into Eq. (S2-1), given by,

$$\frac{\Delta P_3^0}{\Delta u_3^0} = -\frac{\mathbf{i}kL_{333}}{\mu(\omega_{\text{f}}^2 - \omega^2) + G_0 k^2 - \mathbf{i}\gamma\omega}, \quad \text{(S3 - 3)}$$

where $\omega_{\text{f}}$ is the resonant frequency of the fundamental mode ($k$=0) coherent ferron. Substituting Eq.(S3-3) into Eq. (S3-2), one has,

$$c_{33}^{\text{eff}} = c_{33} - \frac{L_{333}^2}{\mu(\omega_{\text{f}}^2 - \omega^2) + G_0 k^2 - \mathbf{i}\gamma\omega}. \quad \text{(S3 - 4)}$$

The real part of the $c_{33}^{\text{tot}}$, Re($c_{33}^{\text{tot}}$), which governs the sound velocity, is given as,

$$\text{Re}(c_{33}^{\text{tot}}) = c_{33} - \frac{L_{333}^2\left[\mu\left(\omega_{\text{f}}^2 - \omega^2\right) + G_0\text{Re}(k^2) - \beta\omega(\text{Im}(k^2) - \gamma\omega)\right]}{[\mu(\omega_{\text{f}}^2 - \omega^2) + G_0\text{Re}(k^2)]^2 + [\text{Im}(k^2) - \gamma\omega]^2}. \quad \text{(S3 - 5)}$$

The longitudinal sound velocity and the corresponding resonant frequency of LA phonons $\omega^{\text{ph}}$ are given by,

$$v_{\text{LA}} = \sqrt{\text{Re}(c_{33}^{\text{tot}})/\rho} \quad \text{and} \quad \omega^{\text{ph}} = v_{\text{LA}}k. \quad \text{(S3 - 6)}$$

We now discus two limiting regimes. In one limiting regime, we consider that the frequency of the polarization oscillation $\omega$ is significantly lower than $\omega_{\text{f}}$ and the wavenumber $k$ is small (i.e., the long-wavelength limit). This case corresponds to, for example, the propagation of a 10-MHz acoustic wave in the experiment reported in [45]. In most ferroelectrics, the condition of $\omega^2 \ll \left(\frac{\mu}{\gamma\beta}\right)\omega_{\mathrm{f}}^2$ would automatically be satisfied if $\omega \ll \omega_{\mathrm{f}}$ based on the typical values of $\gamma$, $\beta$, and $\mu$ (e.g., $\frac{\mu}{\gamma\beta} \sim 1000$ in CIPS). As a result, Eq. (S3-5) reduces to,

$$\mathrm{Re}(c_{33}^{\mathrm{tot}}) \approx c_{33} - \frac{L_{333}^2}{\mu(\omega_{\mathrm{f}}^2)} \equiv c_{33} - \frac{L_{333}^2}{K_{33}} \approx c_{33} - \frac{4{P_3^{\mathrm{eq}}}^2 c_{33}^2 Q_{33}^2}{K_{33}}, \qquad (S3-7)$$

which is equivalent to Eq. (S1-10). In this case, $\mathrm{Re}(c_{33}^{\mathrm{tot}})$ still depends on the temperature and strain because they can modulate $P_3^{\mathrm{eq}}$ and $K_{33}$ (the local curvature), as shown in Figs. 1(d) and 4(a). However, $\mathrm{Re}(c_{33}^{\mathrm{tot}})$ is independent of the frequency of the acoustic wave $\omega$.

In another limiting regime, if we consider a relatively large gradient energy coefficient $G_0$ such that the term $G_0\mathrm{Re}(k^2)$ dominates, Eq. (S3-5) reduces to,

$$\mathrm{Re}(c_{33}^{\mathrm{tot}}) = c_{33} - \frac{L_{333}^2}{G_0\mathrm{Re}(k^2)} \approx c_{33}. \qquad (S3-8)$$

In this case, the sound velocity and the acoustic phonon resonance are only determined by the bare (zero-polarization) elastic stiffness coefficient $c_{33}$, which is the case considered in the main text.

For all the remaining scenarios, $\mathrm{Re}(c_{33}^{\mathrm{tot}})$ strongly depends on the $\omega$, $k$, $\omega_{\mathrm{f}}$, $G_0$, and $P_3^{\mathrm{eq}}$. Since both $\omega_{\mathrm{f}}$ and $P_3^{\mathrm{eq}}$ can be significantly modulated by bias electric field, temperature, and static strain, both the sound velocity and the acoustic phonon resonance can be effectively tuned by these external stimuli, thus opening new opportunities for the discovery of new physical phenomena and device concepts.

## S4. Derivation of the ferron-phonon coupling strength, decoherence rates, and quality factor

*On the ferron-phonon coupling strength*

Applying the partial fraction expansion of the tangent function, Eq.(S2-12b) can be rewritten as,

$$\Omega_{333}(\omega) \approx -\frac{8L_{333}}{d^2\rho}\sum_{n}^{1,3,5\ldots}\left({\omega_n^{\mathrm{ph}}}^2-\omega^2-\mathbf{i}\beta\omega^3\right)^{-1}, \qquad \text{(S4 − 1)}$$

where $\omega_n^{\mathrm{ph}}=\frac{n\pi}{d}\sqrt{\frac{c_{33}}{\rho}}$, and $n$=1,3,5… is odd numbered. Substituting Eq.(S4-1) into Eq.(S2-8), $\chi_{33}(\omega)$ can be written as,

$$\chi_{33}(\omega)=\frac{1}{\mu\kappa_0}\frac{1}{\left(\omega_{\mathrm{f}}^2-\omega^2-\mathbf{i}\frac{\gamma}{\mu}\omega\right)-\frac{8L_{333}^2}{\mu\rho d^2}\sum_n^{1,3,5\ldots}\left({\omega_n^{\mathrm{ph}}}^2-\omega^2-\mathbf{i}\beta\omega^3\right)^{-1}}, \qquad \text{(S4 − 2)}$$

One notes that $\chi_{33}$ is equivalent to the dielectric susceptibility of the following coupled equations with the normalized variables $X=\sqrt{\mu}\Delta P_3(t)$, and $Y_n=\frac{d\sqrt{\rho}}{2\sqrt{2}}\langle\Delta\varepsilon_{33}^n\rangle(t)$. When neglecting the damping coefficient, the coupled equations can be written as,

$$\frac{\partial^2 X}{\partial t^2}+\omega_{\mathrm{f}}^2X+g\sum_n^{1,3,5\ldots}Y_n=0, \qquad \text{(S4 − 3a)}$$

$$\frac{\partial^2 Y_n}{\partial t^2}+{\omega_n^{\mathrm{ph}}}^2Y_n+gX=0, \qquad \text{(S4 − 3b)}$$

where $g=\frac{2\sqrt{2}|L_{333}|}{d\sqrt{\rho\mu}}$. The classical Hamiltonian for such hybrid ferron-phonon system can be written as $\mathcal{H}=T+U$, where the kinetic energy $T$ includes the kinetic energies of the ferron and all acoustic phonon modes, the potential energy $U$ contains the potential energies of uncoupled ferron and phonon subsystems as well as the interaction potential energy. Specifically, one has,

$$\mathcal{H}=\frac{1}{2}\dot{X}^2+\frac{1}{2}\omega_{\mathrm{f}}^2X^2+\sum_n^{1,3,5\ldots}\left(\frac{1}{2}{\dot{Y}_n}^2+\frac{1}{2}{\omega_n^{\mathrm{ph}}}^2Y_n^2\right)+gX\sum_n^{1,3,5\ldots}Y_n \qquad \text{(S4 − 4)}$$

In quantum theory, the coupling strength $g_c$ between two bosonic modes (e.g., phonon, magnon, exciton, ferron) is part of the prefactor in the bilinear interaction term of an operator-form Hamiltonian [59,61],

$$\mathcal{H}_{\mathrm{int}}=\hbar g_c(c^\dagger+c)(d^\dagger+d), \qquad \text{(S4 − 5)}$$

where $(c^\dagger,c)$ and $(d^\dagger,d)$ are annihilation operators. To rewrite the classical Hamiltonian in Eq.(S4-4) into the operator-form Hamiltonian, we introduce the bosonic creation–annihilation operators $(a^\dagger,a)$ for the ferron and $\left(b_n^\dagger,b_n\right)$ for the $n$th acoustic phonon mode, where $[a,a^\dagger]=1$, and $\left[b_n,b_m^\dagger\right]=\delta_{nm}$. The normalized coordinates can be rewritten as,

$$X = \sqrt{\frac{\hbar}{2\omega_{\mathrm{f}}}}(a^{\dagger} + a), \dot{X} = -\mathbf{i}\sqrt{\frac{\hbar\omega_{\mathrm{f}}}{2}}(a^{\dagger} - a), \qquad (\mathrm{S4-6a})$$

$$Y_n = \sqrt{\frac{\hbar}{2\omega_n^{\mathrm{ph}}}}(b_n^{\dagger} + b_n), \dot{Y_n} = -\mathbf{i}\sqrt{\frac{\hbar\omega_n^{\mathrm{ph}}}{2}}(b_n^{\dagger} - b_n), \qquad (\mathrm{S4-6b})$$

which ensure the canonical commutators $[X, \dot{X}] = \mathbf{i}\hbar$, $[Y_n, \dot{Y_m}] = \mathbf{i}\hbar\delta_{nm}$. Substituting Eqs. (S4-6a,b) into Eq.(S4-4), the latter can be rewritten into,

$$\mathcal{H} = \hbar\omega_{\mathrm{f}}\left(a^{\dagger}a + \frac{1}{2}\right) + \sum_{n}^{1,3,5\ldots} \hbar\omega_n^{\mathrm{ph}}\left(b_n^{\dagger}b_n + \frac{1}{2}\right) + \sum_{n}^{1,3,5\ldots} \hbar g_c(a^{\dagger} + a)(b_n^{\dagger} + b_n) \quad (\mathrm{S4-7})$$

where the coupling strength $g_c$ between the ferron and the *n*th acoustic phonon mode is given by,

$$g_c = \frac{g}{2\sqrt{\omega_{\mathrm{f}}\omega_n^{\mathrm{ph}}}} = \frac{\sqrt{2}|L_{333}|}{d\sqrt{\rho\mu\omega_{\mathrm{f}}\omega_n^{\mathrm{ph}}}} \qquad (\mathrm{S4-8})$$

At resonance, $\omega_{\mathrm{f}}$=$\omega_n^{\mathrm{ph}}$=$\omega_0$, Eq. (S4-8) reduces to,

$$g_c = \frac{\sqrt{2}|L_{333}|}{d\omega_0\sqrt{\rho\mu}}. \qquad (\mathrm{S4-9})$$

*On the dissipation rates of uncoupled ferron and phonon modes*

The decoherence rates $\kappa_{\mathrm{f}}$ and $\kappa_{\mathrm{ph}}$ denote the half width at half maximum (HWHM, i.e., the linewidth) of the absorption peak in the power spectra of uncoupled ferrons and acoustic phonons, respectively. They can be derived, respectively, from the dielectric susceptibility of pure ferron system and the mechanical susceptibility of pure phonon system.

The dielectric susceptibility of the pure ferron system is essentially Eq. (S2-8), yet dropping the term $L_{333}\Omega_{333}(\omega)$ from its denominator. Near the ferron resonant frequency, i.e. $\omega \approx \omega_{\mathrm{f}}$, the imaginary part of the dielectric susceptibility, $\mathrm{Im}(\chi_{33})$, can be written as,

$$\mathrm{Im}(\chi_{33}) \approx \frac{\gamma}{4\mu^2\kappa_0\omega_{\mathrm{f}}}\frac{1}{(\omega - \omega_{\mathrm{f}})^2 + \left(\frac{\gamma}{2\mu}\right)^2}, \qquad (\mathrm{S4-10})$$

According to Eq. (S4-10), $\mathrm{Im}(\chi_{33})$ reaches its maximum value at $\omega = \omega_{\mathrm{f}}$, and its half maximum value at $\omega = \omega_{\mathrm{f}} \pm \frac{\gamma}{2\mu}$.Therefore, the HWHM, $\kappa_{\mathrm{f}}$, is identified as,

$$\kappa_{\mathrm{f}} = \frac{\gamma}{2\mu}, \qquad (\mathrm{S4-11})$$

Incorporating the coupling between $k$≠0 mode coherent ferrons (nonuniform polarization wave) and cavity acoustic phonons, the elastodynamic equation for LA phonons can be written as,

$$\rho\frac{\partial^2\Delta u_3(x_3,t)}{\partial t^2} - c_{33}\frac{\partial}{\partial x_3^2}\left(1 + \beta\frac{\partial}{\partial t}\right)\Delta u_3(x_3,t) - L_{333}\frac{\partial}{\partial x_3}\left(1 + \beta\frac{\partial}{\partial t}\right)\Delta P_3(x_3,t) = f_3(t), \quad (\mathrm{S4-12})$$

where $f_3(t)$ denotes volumetric mechanical force (unit: N/m$^3$) applied along the $x_3$ axis. Substituting $\Delta u_3(x_3,t) = \Delta u_3^0(k,\omega)e^{\mathbf{i}(kx_3-\omega t)}$, $\Delta P_3(x_3,t) = \Delta P_3^0(k,\omega)e^{\mathbf{i}(kx_3-\omega t)}$, $f_3(t) = f_3^0(\omega)e^{-\mathbf{i}\omega t}$, and Eq.(S3-3) into Eq. (S4-12), the latter can be rewritten into,

$$-\rho\omega^2\Delta u_3^0(k,\omega) + c_{33}^{\text{tot}}k^2\Delta u_3^0(k,\omega) = f_3^0(\omega), \qquad (\text{S4}-13)$$

Equation (S4-13) allows us to evaluate the mechanical susceptibility $\chi_{33}^{\text{m}}$ for LA phonons,

$$\chi_{33}^{\text{m}} = \frac{\Delta u_3(x_3,t)}{f_3(t)} = \frac{\Delta u_3^0(k,\omega)}{f_3^0(\omega)} = \frac{1}{c_{33}^{\text{tot}}k^2 - \rho\omega^2}, \qquad (\text{S4}-14)$$

For odd-numbered BAW, the wavenumber $k = \frac{n\pi}{d}$, $n$=1,3,5…, and the resonant frequency $\omega_n^{\text{ph}} = \sqrt{\frac{\text{Re}(c_{33}^{\text{tot}})}{\rho}}k$. At the acoustic resonance, $\omega \approx \omega_n^{\text{ph}}$, $\chi_{33}^{\text{m}}$ can be written as

$$\chi_{33}^{\text{m}} \approx \frac{1}{\rho\left({\omega_n^{\text{ph}}}^2 - \omega^2\right) - \mathbf{i}\text{Im}(c_{33}^{\text{tot}})k^2}. \qquad (\text{S4}-15)$$

where the imaginary part of $c_{33}^{\text{tot}}$ , $\text{Im}(c_{33}^{\text{tot}})$, is given by,

$$\text{Im}(c_{33}^{\text{tot}}) \approx -\beta\omega c_{33} + \frac{L_{333}^2\left[\beta\omega\left[\mu\left(\omega_{\text{f}}^2-\omega^2\right)+G_0\text{Re}(k^2)\right]+\text{Im}(k^2)-\gamma\omega\right]}{[\mu(\omega_{\text{f}}^2-\omega^2)+G_0\text{Re}(k^2)]^2+[\text{Im}(k^2)-\gamma\omega]^2}, \qquad (\text{S4}-16)$$

Based on Eq. (S4-15), the imaginary part of $\chi_{33}^{\text{m}}$ reaches its maximum value at $\omega = \omega_n^{\text{ph}}$, and its half maximum at $\omega = \sqrt{{\omega_n^{\text{ph}}}^2 \pm \left|\frac{\text{Im}(c_{33}^{\text{tot}})}{\text{Re}(c_{33}^{\text{tot}})}\right|{\omega_n^{\text{ph}}}^2}$, where $\text{Re}(c_{33}^{\text{tot}})$ and $\text{Im}(c_{33}^{\text{tot}})$ need to be calculated by letting $\omega = \omega_n^{\text{ph}}$. Therefore, the HWHM $\kappa_{\text{ph}}$ can be calculated from the average of these two square root values, i.e.,

$$\kappa_{\text{ph}} \approx \frac{1}{2}\left|\frac{\text{Im}(c_{33}^{\text{tot}})}{\text{Re}(c_{33}^{\text{tot}})}\right|\omega_n^{\text{ph}}, \qquad (\text{S4}-17)$$

When $G_0$ is large, the polarization amplitudes of $k\neq0$ mode coherent ferrons are negligible, one has $\text{Re}(c_{33}^{\text{tot}}) \approx c_{33}$ (see Eq. (S3-8)) and $\text{Im}(c_{33}^{\text{tot}}) = -\beta c_{33}\omega_n^{\text{ph}}$. Eq. (S4-17) reduces to,

$$\kappa_{\text{ph}} = \frac{\beta}{2}{\omega_n^{\text{ph}}}^2. \qquad (\text{S4}-18)$$

The quality $Q$ factor of the phonon subsystem can be calculated as,

$$Q = \frac{\omega_n^{\text{ph}}}{\Delta\omega} = \frac{\omega_n^{\text{ph}}}{2\kappa_{\text{ph}}} \approx \left|\frac{\text{Re}(c_{33}^{\text{tot}})}{\text{Im}(c_{33}^{\text{tot}})}\right|. \qquad (\text{S4}-19)$$

When the polarization amplitudes of $k\neq0$ mode coherent ferrons are negligible, the $Q$ factor reduces to $Q$=1/($\beta\omega_n^{\text{ph}}$), which is independent of temperature.

When the polarization amplitudes of $k\neq0$ mode coherent ferrons are not negligible, e.g., in the case of relatively small $G_0$, both the $\omega_n^{\mathrm{ph}}$ and $Q$ factor become temperature dependent. As examples, Figure S3 shows the temperature dependence of $\omega_{n=1}^{\mathrm{ph}}$ and its corresponding $Q$ factor under $G_0$= $5\times10^{-10}$ J m$^3$/C$^2$ and $d$ = 200 nm, which are calculated based on Eq. (S3-6) and Eq. (S4-19), respectively.

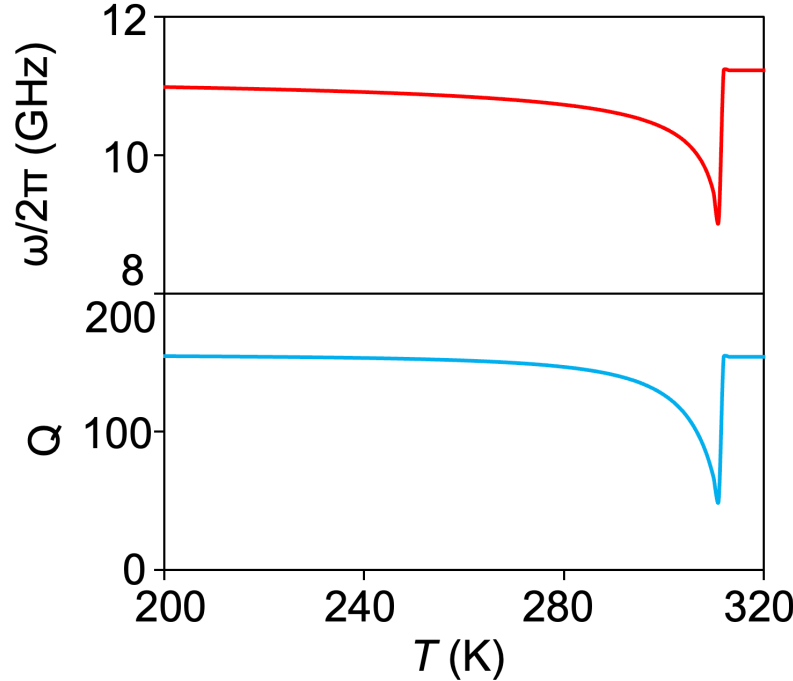

**Figure S2**. Temperature-dependent resonance frequency and quality factor of the standing LA phonons in a CIPS membrane with $G_0 = 5\times10^{-10}$ J m$^3$/C$^2$ and $d = 200$ nm.

## S5. Extraction of the elastic damping coefficient

In [64], the Brillouin light scattering (BLS) spectrum was measured experimentally in the Stokes and anti-Stokes regime of CIPS at room temperature. The relationship between the Brillouin scattering intensity $I(\boldsymbol{q}, \omega)$ and the mechanical displacement susceptibility can be written as [65],

$$I(\boldsymbol{q}, \omega) \propto (n(\omega) + 1)\mathrm{Im}(\chi_{33}^{\mathrm{m}}) \propto \mathrm{Im}(\chi_{33}^{\mathrm{m}}), \qquad (\mathrm{S5}-1)$$

where $n(\omega) = \left(e^{\hbar\omega/k_{\mathrm{B}}T} - 1\right)^{-1}$ denotes the Bose–Einstein occupation factor, $\chi_{33}^{\mathrm{m}}$ denotes the the mechanical susceptibility for the LA phonons.

In this case, the peak frequency in the BLS spectrum corresponds to the resonance frequency of the acoustic phonons, while the linewidth of the peak describes the energy dissipation rate. As discussed in S4, the imaginary part of $\chi_{33}^{\mathrm{m}}$ can be written as,

$$\mathrm{Im}(\chi_{33}^{\mathrm{m}}) \approx \frac{\beta\omega_n^{\mathrm{ph}}}{4\rho} \frac{1}{\left(\omega - \omega_n^{\mathrm{ph}}\right)^2 + \left(\frac{\beta}{2}\omega_n^{\mathrm{ph}^2}\right)^2}, \qquad (\mathrm{S5}-2)$$

In [64], the peak in the BLS spectrum was fitting using the following Lorentzian function,

$$I = I_0 + \frac{(A/\pi)(w/2)}{(f - f_c)^2 + (w/2)^2} = I_0 + \frac{Aw}{(2\pi f - 2\pi f_c)^2 + (2\pi w/2)^2}, \qquad (\mathrm{S5}-3)$$

where $A$ and $w$ are fitting parameters, $f_{\mathrm{c}}$ denotes the center frequency of the scattering spectrum, and $\pi w$ represent the FWHM, with $f_{\mathrm{c}}$=34.40 GHz, $w$=0.684 GHz [64].

Comparing Eq. (S5-2) and Eq. (S5-3), one can see that $\omega_n^{\mathrm{ph}} \equiv 2\pi f_c$ and $\frac{\beta}{2}\omega_n^{\mathrm{ph}^2} \equiv \pi w$, yielding,

$$\beta = \frac{w}{2\pi f_c^2} = 9.19 \times 10^{-14}\ \mathrm{s}, \qquad (\mathrm{S5}-4)$$

### S6. Time-domain solutions of $\Delta P_3(t)$ and $\Delta\varepsilon_{33}(x_3,t)$

The time-domain oscillation of polarization $\Delta P_3(t)$ can be calculated as,

$$\Delta P_3(t)=\kappa_0\chi_{33}(t)*E_3^{\text{inc}}(t)=\kappa_0\int_0^{\infty}\chi_{33}(\tau)E_3^{\text{inc}}(t-\tau)\,d\tau, \qquad (\text{S6}-1)$$

where $\chi_{33}(t)$ can be obtained by performing inverse Fourier transform of $\chi_{33}(\omega)$ (Eq. (S2-7)). When the frequency of the driving microwave field is close to the resonance frequency of an odd-numbered cavity bulk acoustic phonon, i.e., $\omega\approx\omega_n^{\text{ph}}=\frac{n\pi}{d}\sqrt{\frac{c_{33}}{\rho}}$, $n$ =1,3,5,…, the spatial distribution of strain takes the form of $\Delta\varepsilon_{33}(x_3,\omega)=\frac{n\pi}{2}\langle\Delta\varepsilon_{33}\rangle(\omega)\sin\left(\frac{n\pi}{d}x_3\right)$, where $\langle\Delta\varepsilon_{33}\rangle(\omega)$=$\Omega_{333}(\omega)\Delta P_3$ and the expression of $\Omega_{333}(\omega)$ is provided in Eq. (S2-12b). The time-domain oscillation of strain $\Delta\varepsilon_{33}(x_3,t)$ can then be obtained by inverse Fourier transform.

The detailed analytical expressions of $\Delta P_3(t)$ and $\Delta\varepsilon_{33}(x_3,t)$ are cumbersome, and thus omitted herein. However, it is worth noting that the attenuation of both $\Delta P_3(t)$ and $\Delta\varepsilon_{33}(x_3,t)$ is governed by an identical damping parameter, i.e., $\lambda=\frac{\gamma+\beta\mu\omega_0^2}{4\mu}$. Moreover, the envelope of $\Delta P_3(t)$ and $\Delta\varepsilon_{33}(x_3,t)$ can be expressed as,

$$\Delta P_3^{\text{env}}(t)=\Delta P_3^{\text{env},0}e^{-\lambda t}, \qquad (\text{S6}-2\text{a})$$

$$\Delta\varepsilon_{33}^{\text{env}}(x_3,t)=\Delta\varepsilon_{33}^{\text{env},0}e^{-\lambda t}\frac{n\pi}{2}\sin\left(\frac{n\pi}{d}x_3\right), \qquad (\text{S6}-2\text{b})$$

where $\Delta P_3^{\text{env},0}$ and $\Delta\varepsilon_{33}^{\text{env},0}$ are the largest peak amplitude of $\Delta P_3(t)$ and $\Delta\varepsilon_{33}(x_3,t)$, respectively.

As one example, we calculate $\Delta P_3(t)$ and $\Delta\varepsilon_{33}$($x_3$=$d$/2,$t$) in response to a Gaussian-enveloped electric field pulse $E_3^{\text{inc}}(t)$ with a center temporal frequency of 46 GHz, as shown in the top panel of Fig. S3(a). Specifically, one has $E_3^{\text{inc}}(t)=E_3^0e^{-(t-5\tau)^2/2\tau^2}\cos\big(\omega_0(t-5\tau)\big)$, where $E_3^0$=0.1 MV/m, $\tau$=10 ps, and $\omega_0$=46 GHz. Comparing the two panels in Figs. S3(a), one can see the onset of coherent beating oscillation approximately after 58 ps, where the maxima of $\Delta P_3(t)$ correspond to an almost zero $\Delta\varepsilon_{33}(x_3,t)$. The attenuation of the peak amplitudes in the temporal profiles can be well fitted by $\Delta P_3(t)=\Delta P_3^0e^{-\lambda t}$ and $\Delta\varepsilon_{33}(x_3,t)$=$\Delta\varepsilon_{33}^0e^{-\lambda t}$ (see green dashed lines in both panels of Fig. S3(a)), where $\Delta P_3^0$ and $\Delta\varepsilon_{33}^0$ are the amplitude of the highest peak.

The validity and accuracy of the analytical solutions are further corroborated by their good agreement with the numerical simulation results from dynamical phase-field model (DPFM), which relies on the numerical solutions of the coupled equations of motion for $P_i$ and $u_i$ [see Eqs. (S2-1,2)]. In the present dynamical phase-field simulations, a 1D discretized system with a total thickness of 48.9 nm was constructed along the $x_3$ axis to represent the freestanding CIPS membrane and the air layers. The cell size is 0.1 nm. The CIPS membrane occupies the cells from 11 to 499. Two 1 nm-thick air layers are added on both the top and bottom of the CIPS film. Periodic boundary conditions were applied along the $x_1$-$x_2$ plane. The temperature is set to 298 K. The numerical solutions are obtained using a time step of $\Delta t=2\times10^{-15}$ s. Regarding the simulation data in Fig. S4, $\Delta P_3(t)$ was extracted as the spatial average along the entire thickness of the CIPS, i.e., $\Delta P_3(t)$=$\langle\Delta P_3\rangle(t)$; $\Delta\varepsilon_{33}(t)$ was extracted based on the local data in the middle of the CIPS, i.e., $\Delta\varepsilon_{33}(t)$= $\Delta\varepsilon_{33}$($x_3$=34.5 nm, $t$). The applied electric field $E_3^{\mathrm{inc}}(t)$ is identical to that used in analytical calculation. The isotropic gradient energy coefficient $G_0$ is set to 5.1×10$^{-6}$ J m$^3$/C$^2$, which is sufficiently large to suppress $k$≠0 mode coherent ferron, ensuring that the polarization oscillation is largely spatially uniform (see Fig. S3(b)). More details of our dynamical phase-field model, including the governing equations and numerical implementation, can be found in our previous work [30-32,46]. The coherent beating oscillation shown in Fig. S3(a) suggests a coherent and complete dynamical energy exchange between the ferron and phonon systems (also see Fig. 1(f) in the main text), which is a canonical time-domain feature of strong coupling in a two-level system. Accordingly, the corresponding frequency spectrum exhibits a clear splitting into two distinct peaks, as shown in Fig.S3(c). The splitted peaks appear at 37.5 and 52.5 rad GHz, and the half of the frequency gap $\Delta\omega/2\pi$=7.5 GHz agrees well with the coupling strength $g_c/2\pi$=7.27 GHz in Eq. (2) of the main text.

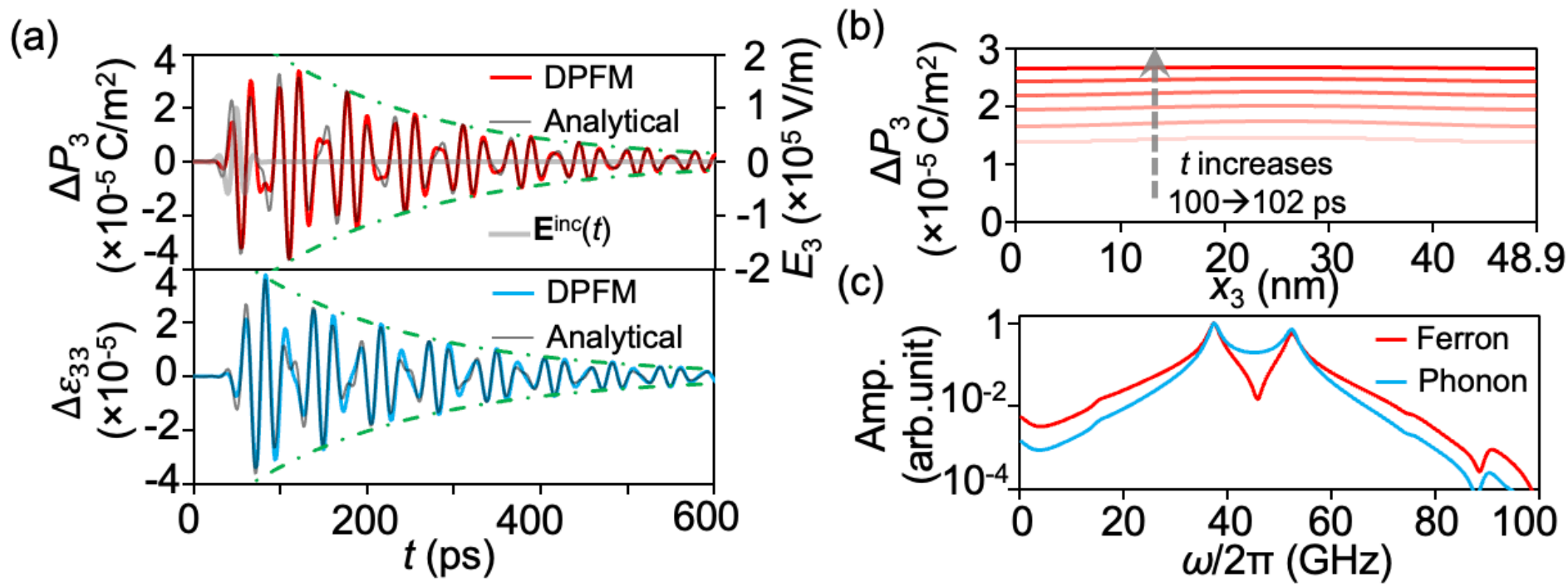

**Figure S3**. (**a**) Temporal evolution of spatially-averaged dynamic polarization $\langle\Delta P_3\rangle$ ($t$)= $\frac{\int P_3(t)dx_3}{d} - P_3(t=0)$, and dynamic strain $\Delta\varepsilon_{33}$ ($x_3$=$d$/2,$t$)= $\varepsilon_{33}$ ($x_3$=$d$/2,$t$)- $\varepsilon_{33}$ ($x_3$=$d$/2,$t$=0) of the ferron and phonon subsystems in a 48.9-nm-thick CIPS membrane under the excitation by a Gaussian-enveloped electric field pulse $E_3^{\mathrm{inc}}(t)$. The profile of the latter is also shown in (**a**). (**b**) Spatially profiles of the local polarization along the thickness direction $\Delta P_3$($x_3$,$t$)=$P_3$($x_3$,$t$)-$P_3$($t$=0) from 100 ps to 102 ps. (**c**) Frequency spectra of ferron and phonon systems, obtained by the Fourier transform of the time-domain data of polarization and strain oscillation in (**a**). In (**b,c**), the results are obtained from dynamical phase-field simulations (DPFM). The polarization damping $\gamma$=10$^{-3}$ Ω·m and the elastic damping $\beta$=9.19×10$^{-14}$ s. The temperature is 298 K.

## S7. Energies of the ferron and acoustic phonon systems

*Temporal evolution of the energy densities in the ferron system*

Let us first consider the energies of the ferron system. For the single-domain CIPS hosting only *k*=0 mode coherent ferron, Eq. (S2-1) can be rewritten into,

$$\mu\frac{\partial^2 P_3}{\partial t^2}+\gamma\frac{\partial P_3}{\partial t}+\frac{\partial f^{\text{Landau}}}{\partial P_3}=-\left(\frac{\partial f^{\text{Elec}}}{\partial P_3}+\frac{\partial f^{\text{Elast}}}{\partial P_3}\right), \qquad (S7-1)$$

By integrating both sides of the Eq.(S7-1) from $P_3^{\text{eq}}$ (*t*=0) to $P_3(t)$ and averaging over the thickness, one can write down the energy conservation relation,

$$T^{\text{f}}+Q_{\text{dis}}^{\text{f}}+\Delta U_{\text{Landau}}^{\text{f}}=\Delta U_{\text{Elec}}^{\text{f}}+\Delta U_{\text{elast}}^{\text{f}}, \qquad (S7-2)$$

where $T^{\text{f}}$ is the instantaneous kinetic energy density of the ferron system, given by,

$$T^{\text{f}}=\frac{1}{d}\int_0^d\int_{P_3^0}^{P_3(t)}\left(\mu\frac{\partial^2 P_3}{\partial t^2}\right)dP_3\,dx_3=\frac{1}{2}\mu\left(\frac{\partial P_3}{\partial t}\right)^2, \qquad (S7-3)$$

$Q_{\text{dis}}^{\text{f}}$ is the energy density dissipated from the ferron system at a given moment *t*, given by,

$$Q_{\text{dis}}^{\text{f}}=\frac{1}{d}\int_0^d\int_{P_3^0}^{P_3(t)}\left(\gamma\frac{\partial P_3}{\partial t}\right)dP_3\,dx_3=\int_0^t\left(\gamma\left(\frac{\partial P_3}{\partial t}\right)^2\right)dt, \qquad (S7-4)$$

$\Delta U_{\text{Landau}}^{\text{f}}$ is the instantaneous Landau free energy density, given by,

$$\Delta U_{\text{Landau}}^{\text{f}}=\frac{1}{d}\int_0^d\int_{P_3^{\text{eq}}}^{P_3(t)}\left(\frac{\partial f^{\text{Landau}}}{\partial P_3}\right)dP_3\,dx_3=f^{\text{Landau}}\big(P_3(t)\big)-f^{\text{Landau}}\big(P_3^{\text{eq}}\big), \qquad (S7-5)$$

Under zero depolarization field ($E_i^{\text{d}}=0$) and zero radiation electric field ($E_i^{\text{rad}}=0$), $\Delta U_{\text{Elec}}^{\text{f}}$ is the amount of electric work done by the incident electric field, $E_3^{\text{inc}}$, to the ferron system, given by,

$$\Delta U_{\text{Elec}}^{\text{f}}=\frac{1}{d}\int_0^d\int_{P_3^{\text{eq}}}^{P_3(t)}\left(-\frac{\partial f^{\text{Elec}}}{\partial P_3}\right)dP_3\,dx_3=\int_{P_3^{\text{eq}}}^{P_3(t)}E_3^{\text{inc}}(t)dP_3, \qquad (S7-6)$$

$\Delta U_{\text{elast}}^{\text{f}}$ is the instantaneous elastic energy density of the ferron system, given by,

$$\Delta U_{\text{elast}}^{\text{f}}=-\frac{1}{d}\int_0^d\int_{P_3^{\text{eq}}}^{P_3(t)}\left(\frac{\partial f^{\text{Elast}}}{\partial P_3}\right)dP_3\,dx_3. \qquad (S7-7)$$

The expression of $f^{\text{Elast}}$ is provided in Supplemental section S1. After some algebra, we found that $\Delta U_{\text{elast}}^{\text{f}}$ can be separated into two parts, including (i) an energy density that is determined by $P_3$ and $\varepsilon_{33}$ at the initial (*t*=0) state and the moment *t*, denoted as $\Delta U_{\text{elast,1}}^{\text{f}}$, and (ii) a term that depends on the evolution history of $P_3$ and $\varepsilon_{33}$, denoted as $\Delta U_{\text{elast,2}}^{\text{f}}$. Specifically,

$$\Delta U^{\mathrm{f}}_{\mathrm{elast,1}} = (A_1 Q_{13} + A_2 Q_{23} + A_4 Q_{53}) {P_3^{\mathrm{eq}}}^2 P_3^2 |_0^t - \frac{1}{2} A_5 P_3^4 |_0^t + \frac{1}{d} \int_0^d (A_3 P_3^2 \varepsilon_{33} |_0^t) dx_3 \quad \text{(S7 − 8)}$$

where the coefficients $A_1 = c_{11}Q_{13} + c_{12}Q_{23} + c_{13}Q_{33} + 2c_{15}Q_{53}$ , $A_2 = c_{12}Q_{13} + c_{22}Q_{23} + c_{23}Q_{33} + 2c_{25}Q_{53}$ , $A_3 = c_{13}Q_{13} + c_{23}Q_{23} + c_{33}Q_{33} + 2c_{35}Q_{53}$ , $A_4 = c_{15}Q_{13} + c_{25}Q_{23} + c_{35}Q_{33} + 2c_{55}Q_{53}$ , $A_5 = c_{11}Q_{13}^2 + 2c_{12}Q_{13}Q_{23} + c_{22}C_{23}^2 + 2c_{13}Q_{13}Q_{33} + 2c_{23}Q_{23}Q_{33} + c_{33}Q_{33}^2 + 4c_{15}Q_{13}Q_{53} + 4c_{25}Q_{23}Q_{53} + 4c_{35}Q_{33}Q_{53} + 4c_{55}Q_{53}^2$.

The evolution-history dependent term is given as,

$$\Delta U^{\mathrm{f}}_{\mathrm{elast,2}} = -A_3 \int_0^t \left( P_3^2 \frac{\partial \langle \Delta\varepsilon_{33} \rangle}{\partial t} \right) dt \approx -c_{33} Q_{33} \int_0^t \left( P_3^2 \frac{\partial \langle \Delta\varepsilon_{33} \rangle}{\partial t} \right) dt, \qquad \text{(S7 − 9)}$$

which is relevant to the mechanical work done by the phonon system to the ferron system.

*Temporal evolution of the energy densities in the phonon system*

Next, we evaluate the energies of the bulk cavity acoustic phonons. For the single-domain CIPS, the equation of motion for LA phonons $\Delta u_3(x_3, t)$ is given as,

$$\rho \frac{\partial^2 u_3(x_3,t)}{\partial t^2} - \beta \frac{\partial^2 \sigma_{33}(x_3,t)}{\partial t\, \partial x_3} = \frac{\partial \sigma_{33}(x_3,t)}{\partial x_3}, \qquad \text{(S7 − 10)}$$

By integrating both side of Eq.(S7-10) from $u_3^{\mathrm{eq}}$ to $u_3(t)$ and averaging over the thickness, one can write down the energy conservation relation,

$$T^{\mathrm{ph}} + Q^{\mathrm{ph}}_{\mathrm{dis}} = \Delta U^{\mathrm{ph}}_{\mathrm{elast}}, \qquad \text{(S7 − 11)}$$

where $T^{\mathrm{ph}}$ is the instantaneous kinetic energy density of the acoustic phonons, given by,

$$T^{\mathrm{ph}} = \frac{1}{d} \int_0^d \int_{u_3^{\mathrm{eq}}}^{u_3(t)} \left( \rho \frac{\partial^2 u_3(x_3,t)}{\partial t^2} \right) du_3\, dx_3 = \frac{\rho}{2d} \int_0^d \left( \frac{\partial u_3}{\partial t} \right)^2 dx_3. \qquad \text{(S7 − 12)}$$

$Q^{\mathrm{f}}_{\mathrm{dis}}$ is the energy density that has been dissipated from the phonon system at a given time moment $t$, given by,

$$Q^{\mathrm{ph}}_{\mathrm{dis}} = \frac{1}{d} \int_0^d \int_{u_3^{\mathrm{eq}}}^{u_3(t)} \left( -\beta \frac{\partial^2 \sigma_{33}(x_3,t)}{\partial t\, \partial x_3} \right) du_3\, dx_3. \qquad \text{(S7 − 13)}$$

Applying the traction-free boundary condition at the top and bottom surface of the CIPS membrane, i.e. $\sigma_{33}(x_3 = d, t) = \sigma_{33}(x_3 = 0, t) = 0$, Eq. (S7-13) can be written as,

$$Q^{\mathrm{ph}}_{\mathrm{dis}} = \beta \frac{c_{33}}{d} \int_0^d \int_0^t \left( \left( \frac{\partial \varepsilon_{33}}{\partial t} \right)^2 \right) dt\, dx_3 - \beta \frac{1}{d} A_3 \int_0^d \int_0^t \left( \frac{\partial (P_3^2)}{\partial t} \frac{\partial \varepsilon_{33}}{\partial t} \right) dt\, dx_3, \quad \text{(S7 − 14)}$$

$\Delta U^{\mathrm{ph}}_{\mathrm{elast}}$ is the instantaneous elastic energy density of the acoustic phonons, given by,

$$\Delta U_{\text{elast}}^{\text{ph}} = \frac{1}{d}\int_0^d \int_{u_3^{\text{eq}}}^{u_3(t)} \left(\frac{\partial}{\partial x_3}\big(\sigma_{33}(x_3,t)\big)\right) du_3\, dx_3\,, \qquad \text{(S7 − 15)}$$

$\Delta U_{\text{elast}}^{\text{ph}}$ can be separated into three parts, including (i) an energy density that is determined by $\varepsilon_{33}$ at the initial ($t$=0) state and the moment $t$ and coupled directly to $P_3^{\text{eq}}$, i.e.,

$$\Delta U_{\text{elast,1}}^{\text{ph}} = -A_3 P_3^{\text{eq}\,2} \frac{1}{d}\int_0^d (\varepsilon_{33}(t) - \varepsilon_{33}(t=0))dx_3\,; \qquad \text{(S7 − 16)}$$

(ii) an energy density that is determined by $\Delta\varepsilon_{33}(x_3,t)$ at the initial ($t$=0) state and the moment $t$ yet does not involve direct coupling to polarization, i.e.,

$$\Delta U_{\text{elast,2}}^{\text{ph}} = -\frac{c_{33}}{2}\frac{1}{d}\int_0^d (\Delta\varepsilon_{33}^2(t) - \Delta\varepsilon_{33}^2(t=0))dx_3\,; \qquad \text{(S7 − 17)}$$

and (iii) a term that depends on the evolution history of $P_3$ and $\varepsilon_{33}$, i.e.,

$$\Delta U_{\text{elast,3}}^{\text{ph}} = A_3 \int_0^t \left(P_3^2 \frac{\partial\langle\Delta\varepsilon_{33}\rangle}{\partial t}\right) dt \approx c_{33}Q_{33}\int_0^t \left(P_3^2 \frac{\partial\langle\Delta\varepsilon_{33}\rangle}{\partial t}\right) dt. \qquad \text{(S7 − 18)}$$

Comparing Eq. (S7-18) and Eq. (S7-9), one can see that $\Delta U_{\text{elast,3}}^{\text{ph}} = -\Delta U_{\text{elast,2}}^{\text{f}}$. $\Delta U_{\text{elast,3}}^{\text{ph}}$ is therefore relevant to the mechanical work done by the ferron system to phonon system. Furthermore, one can now rewrite Eq. (S7-11) into,

$$T^{\text{ph}} + Q_{\text{dis}}^{\text{ph}} - \Delta U_{\text{elast,2}}^{\text{ph}} = \Delta U_{\text{elast,1}}^{\text{ph}} + \Delta U_{\text{elast,3}}^{\text{ph}}. \qquad \text{(S7 − 19)}$$

We further define the intrinsic energy of each system as the sum of its kinetic energy density and its intrinsic potential energy, which refers to the energy densities that do not involve direct coupling to the other system. In this regard, the intrinsic energy of the ferron system can be written as,

$$f^{\text{ferron}} = T^{\text{f}} + \Delta U_{\text{Landau}}^{\text{f}}, \qquad \text{(S7 − 20)}$$

where $\Delta U_{\text{Landau}}^{\text{f}}$ does not contain terms that are coupled to strain. Likewise, the intrinsic energy of the phonon system can be written as

$$f^{\text{phonon}} = T^{\text{ph}} - \Delta U_{\text{elast,2}}^{\text{ph}}, \qquad \text{(S7 − 21)}$$

where $-\Delta U_{\text{elast,2}}^{\text{ph}}$ does not contain terms that are coupled to polarization.

### S8. Dynamic phase-field simulation of the strong coupling between ferrons and cavity acoustic phonons with a relatively small gradient energy coefficient $G_0$.

We used a dynamical phase-field model (DPFM) to simulate the spatially averaged polarization $\langle\Delta P_3\rangle(t)$ and the strain in the middle plane of the CIPS membrane, $\Delta\varepsilon_{33}(x_3{=}d/2,t)$, under the same excitation electric-field pulse as in Fig. S3 yet a much smaller gradient energy coefficient $G_0$. The temporal profiles of $\langle\Delta P_3\rangle(t)$ and $\Delta\varepsilon_{33}(x_3{=}d/2,t)$ are shown in Fig. S4(a). The spatial profiles of $\Delta P_3(x_3,t)$ shown in Fig. S4(b) are non-uniform and contain a DC component. Such nonuniform polarization wave ($k{\neq}0$ coherent ferrons) is induced by the standing acoustic wave $\Delta\varepsilon_{33}(x_3,t)$ ($k{\neq}0$ cavity acoustic phonons) via the piezoelectric coupling. The temporal profiles shown in Fig. S4(a) display a clear feature of coherent beating oscillation after 120 ps, indicating a strong coupling between coherent ferrons and cavity acoustic phonons. However, the frequency spectrum of the ferrons, as shown Fig. S4(c), still shows a peak at $\omega/2\pi{=}\ \omega_f/2\pi = 46$ GHz, in addition to the splitted frequency at 37 GHz and 53.5 GHz, respectively. We believe this spectral feature is an indicator of a tripartite coupling among $k$=0 mode ferrons, $k{=}\pi/d$ ($n$=1) mode ferrons (with a resonant frequency almost identical to $k$=0 ferron due to the relatively small $G_0$, see Fig. 1(c) in the main text), and $k{=}\pi/d$ (n=1) mode acoustic phonons. Detailed analyses of such more complex ferron-phonon coupling are reserved for future studies.

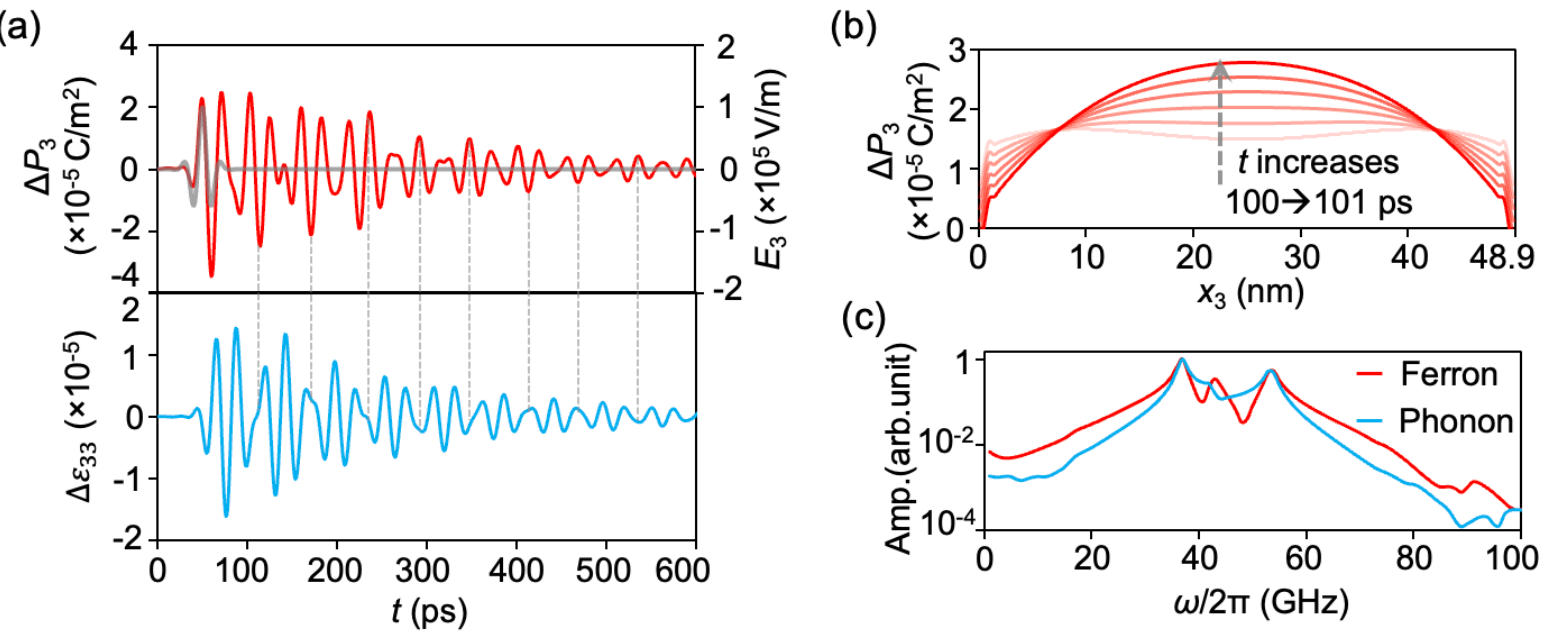

**Figure S4**. (**a**) Temporal evolution of spatially-averaged dynamic polarization $\langle\Delta P_3\rangle\,(t){=}\frac{\int P_3(t)dx_3}{d} - P_3(t=0)$, and dynamic strain $\Delta\varepsilon_{33}(x_3{=}d/2,t){=}\varepsilon_{33}(x_3{=}d/2,t){-}\varepsilon_{33}(x_3{=}d/2,t{=}0)$ under the excitation by a Gaussian-enveloped electric field pulse $E_3^{\mathrm{inc}}(t)$. The profile of the latter is also shown in (**a**). The gray dashed lines indicate the coincidence between the peaks (valleys) of the ferron profile and the valleys (peaks) of the acoustic phonon profile, respectively, a typical feature of beating oscillation. (**b**) Spatially profiles of the local polarization along the thickness direction $\Delta P_3(x_3,t){=}P_3(x_3,t){-}P_3(t{=}0)$ from 100 ps to 102 ps. (**c**) Frequency spectra of ferron and phonon subsystems, obtained by Fourier transform of the time-domain data of polarization and strain oscillation in (**a**). The system setting is the same with Fig. S3 except that a smaller gradient coefficient $G_0{=}5.1{\times}10^{-10}$ J m$^3$/C$^2$ is used.

**Supplemental References**